\documentclass[12pt,preprint]{aastex}

\newcommand{\kev}{keV}
\newcommand{\fe}{Fe~K$\alpha$}
\newcommand{\etal}{et al.}
\newcommand{\oeight}{\ion{O}{8}}
\newcommand{\mcg}{MCG--6-30-15}

\usepackage{emulateapj5}
\usepackage{times}
\usepackage{graphicx}

\begin{document}

\title{X-ray Reflection from Inhomogeneous Accretion Disks: I. Toy
  Models and Photon Bubbles}


\author{D. R. Ballantyne\altaffilmark{1,2},
  N. J. Turner\altaffilmark{3} and O. M. Blaes\altaffilmark{3}}
\altaffiltext{1}{Canadian Institute for Theoretical Astrophysics,
McLennan Labs, 60 St. George Street, Toronto, Ontario, Canada M5S 3H8;
ballantyne@cita.utoronto.ca} 
\altaffiltext{2}{Kavli Institute for
Theoretical Physics, Kohn Hall, University of California, Santa
Barbara, CA 93106}
\altaffiltext{3}{Department of Physics, University of California,
  Santa Barbara, CA 93106; neal, blaes@physics.ucsb.edu}

\begin{abstract}
Numerical simulations of the interiors of radiation dominated
accretion disks show that significant density inhomogeneities can be
generated in the gas. Here, we present the first results of our study
on X-ray reflection spectra from such heterogeneous density
structures. We consider two cases: first, we produce a number of toy
models where a sharp increase or decrease in density of variable width
is placed at different depths in a uniform slab. Comparing the
resulting reflection spectra to those from an unaltered slab shows
that the inhomogeneity can affect the emission features, in particular
the \fe\ and \oeight\ Ly$\alpha$ lines. The magnitude of any
differences depends on both the parameters of the density change and
the ionizing power of the illuminating radiation, but the
inhomogeneity is required to be within $\sim2$ Thomson depths of the
surface to cause an effect. However, only relatively small variations
in density (on the order of a few) are necessary for significant
changes in the reflection features to be possible. Our second test was
to compute reflection spectra from the density structure predicted by
a simulation of the non-linear outcome of the photon bubble
instability. The resulting spectra also exhibited differences from the
constant density models, caused primarily by a strong 6.7~\kev\ iron
line. Nevertheless, constant density models can provide a good fit to
simulated spectra, albeit with a low reflection fraction, between 2
and 10~\kev. Below 2~keV, differences in the predicted soft X-ray line
emission result in very poor fits with a constant density ionized disk
model. The results indicate that density inhomogeneities may further
complicate the relationship between the \fe\ equivalent width and the
X-ray continuum. Calculations are still needed to verify that density
variations of sufficient magnitude will occur within a few Thomson
depths of the disk photosphere.
\end{abstract}


\section{Introduction}
\label{sect:intro}
The reprocessing of hard X-rays in relatively cold, optically thick
material is a major component of the current X-ray phenomenology of
active galactic nuclei (AGN) and Galactic black hole candidates
(GBHCs). The evidence for this is based on the shape of the observed
spectral continuum. Starting with \textit{Ginga} observations in the
late 1980s and continuing on with data collected from \textit{ASCA},
\textit{RXTE} and \textit{Beppo}SAX, it was found that the continuum
of almost all type 1 Seyfert galaxies exhibit a spectral hardening
beyond $\sim$10~\kev\ \citep{pou90,np94} before rolling over at
energies $\ga 100$~\kev\ \citep{per02,ris02}. This
fact combined with the near ubiquitous presence of a \fe\ line
\citep{np94,rey97} is completely consistent with the predictions of
Compton reflection of X-rays in dense, optically thick material
\citep*{gf91,mpp91}. Moreover, the equivalent width (EW) of the \fe\
lines are usually large enough (EW $\sim$ 50--300~eV) that the
reflector must cover about half the sky as seen from the X-ray source
\citep{gf91}, but remain out of the line of sight (in order to not
produce significant absorption). In the complex environment of
the central engine of an AGN, there are many potential locations for
the reprocessing gas, such as broad-line region clouds, and the
obscuring material of the unification schemes. Indeed, reflection from
one or both of these sites may be a common occurrence in Seyfert
galaxies \citep*[e.g.][]{yaq01,kas01,py03}. However, in at least some
objects (most famously, \mcg; e.g., \citealt{fab02}) it was found that
the \fe\ line exhibits a broad red-wing, consistent with originating
from the inner regions of the accretion disk
\citep*{fab89,fab00,rn03}. Thus, in these sources, a detailed
comparison of the X-ray spectrum with models of Compton reflection may
yield information on the structure, metallicity, and ionization state
of the accretion disk at small Schwarzschild radii ($R_{\mathrm{S}} =
2 GM/c^2$, where $M$ is the black hole mass). Evidence for broad \fe\
lines in GBHCs is becoming more common
\citep*[e.g.,][]{mar02,mil02a,mil02b,mil02c,mil03}, so the possibility
of performing similar analyses also exist for the Galactic accreting
black hole systems.

Calculations of X-ray reflection from optically-thick material have
increased in sophistication over the last decade. The initial
Monte-Carlo calculations by \citet{gf91} and \citet{mpp91} assumed a
neutral, constant density slab, and were able to quantify the observed
strength of the \fe\ fluorescence line as a function of viewing angle
and abundance. Soon after, \citet{ros93} and \citet{zyc94} allowed the
gas to be ionized by the incoming X-rays and produced reflection
spectra that included recombination lines and edges as well as
fluorescence lines. \citet{mfr93,mfr96} studied the effects of
ionization on the \fe\ line, while \citet{ros99} clearly showed that
the entire reflection continuum changed shape as a function of the
ionization parameter ($\xi = 4\pi F_{\mathrm{X}} / n_{\mathrm{H}}$,
where $F_{\mathrm{X}}$ is the incident X-ray flux, and
$n_{\mathrm{H}}$ is the density of the reprocessor). More recently,
the assumption of constant density slabs has been replaced with more
complicated density distributions, such as hydrostatic balance
\citep*{nkk00,nk01,brf01,roz02} or constant pressure
\citep{dum02}. The major difference between these variable density
models and the constant density ones is that under certain conditions
(when the Compton temperature of the radiation field is very high) the
illuminated atmosphere can be subject to a thermal instability which
can split the gas into an outer, low density, completely ionized zone
and a deeper, high density, completely recombined zone
\citep*[cf.,][]{kmt81}. When this instability occurs the reflection
spectrum exhibits only neutral features formed in the lower layer that
are slightly broadened due to Compton scattering by the hot surface
layer. Otherwise, the reflection spectra looks much like a diluted
version of an ionized constant density model \citep{brf01,dn01}. The
features are weaker due to both Compton broadening by the hot surface
gas, and the lower emissivity (proportional to density) of the
emitting material.

Over the last few years it has been possible to fit models which
self-consistently predict both the reflection continuum and line
emission to real X-ray data. The constant density models of
\citet{ros93} have been applied to both narrow-line and broad-line
Seyfert~1 galaxies \citep*{bif01,orr01,der02,lon03} as well as GBHCs
\citep{mil03}. While these are not as sophisticated as the hydrostatic
models, they are relatively quick to compute and provide a good
parameterization of the data which allows easy comparison between
objects. On the whole, when combined with a relativistic blurring
function, the models do provide a reasonable description of the hard
X-ray spectrum for many accreting black holes. The exceptions to this,
such as the narrow-line Seyfert~1 1H~0707-495 \citep{bol02} and \mcg\
\citep*{bal03} may indicate that more complicated geometries are
required than a simple flat accretion disk.

As interesting as these fits are, a lingering question remains: are
the constant density or hydrostatic assumptions representative of real
accretion disks? Where radiation pressure exceeds gas pressure, as
likely occurs in the inner regions of AGN disks accreting at more than
0.3\% of the Eddington rate \citep{ss73}, models with stress assumed
proportional to total pressure are viscously \citep{le74}, thermally
\citep{ss76}, and convectively \citep{bkb77} unstable.  These
instabilities may lead to space or time variation in disk structure.
Furthermore, the stresses driving accretion in ionized,
differentially-rotating disks are now thought to be due to magnetic
forces in turbulence resulting from the magneto-rotational instability or
MRI \citep{bh91}. In MHD simulations, gas motions in the turbulence vary
over an orbital period \citep{bh98}, which is similar to the timescale
needed for establishing vertical hydrostatic balance \citep*{fkr02},
Therefore, hydrostatic reflection models may not accurately represent
the structure of turbulent accretion disks. Simulations of the
turbulence which include effects of radiation diffusion indicate that
density fluctuations may exceed an order of magnitude
\citep{tsks03}. Radiation-supported disks with a magnetic field are
subject to another dynamical instability, the photon bubble
instability or PBI \citep{ar92,gam98}. Photon bubble modes having
wavelengths shorter than the gas pressure scale height can grow faster
than the orbital frequency \citep{bs01,bs03}.  The non-linear
development of the instability may lead to trains of shocks
propagating through the disk surface layers \citep{beg01}, leading to
a way of producing super-Eddington luminosities \citep{beg02}.
Density contrasts between shocked and inter-shock regions can exceed
one hundred (N.J. Turner \etal\ 2004, in preparation).  Will density
fluctuations caused by the MRI and PBI affect the reflection spectrum?
\citet{fab202} and \citet*{rfb02} suggested that they may strengthen
the emission features and thus produce an apparent large reflection
fraction. But several unknowns remain: since the reflection features
are formed within 10 Thomson depths ($\tau_{\mathrm{T}}$) of the
surface, what is the magnitude of any density change and where must it
lie in the illuminated gas in order to alter the observed reflection
spectrum? Do the accretion disk simulations produce the necessary
fluctuations? The answers to these questions are crucial to the
interpretation of X-ray spectra of accreting black holes.

In this work, we investigate the properties of X-ray reflection spectra
from heterogeneous atmospheres. This paper (Paper 1) first
considers a toy model where a density cut of variable depth and width
has been sliced out of a constant density slab (\S~\ref{sect:toy}). This will
allow a systematic investigation of the effects of a simple density
inhomogeneity. We will then consider a more complicated structure
produced by calculations of the photon-bubble instability in a
radiation-dominated atmosphere with no turbulence
(\S~\ref{sect:bubbles}). Reflection spectra calculated from all these
structures will then be compared with the standard constant density
ones. In \S~\ref{sect:discuss} we discuss the results from this paper
and form our conclusions in \S~\ref{sect:concl}. In a companion paper
(Paper 2), we apply the knowledge gained here to reflection by
simulated radiation dominated accretion disks.

\section{Toy Model of Inhomogeneous Reflection}
\label{sect:toy}
The simplest test to see if density inhomogeneities affect reflection
spectra is to insert a discontinuous jump or drop in density into an
otherwise constant density model. In
this section, we experiment with placing such steps with different
widths at various places in a uniform slab.

\subsection{Model Setup and Assumptions}
\label{sub:setup}
The calculations used the reflection code described by \citet{ros93}
and \citet{ros99}, and the
reader is referred to those papers for the details on the
computational procedures. The models are computed by solving the
coupled equations of radiative transfer and thermal and ionisation
balance in one spatial dimension, using the two-stream approximation
for the incoming radiation and a Fokker-Planck/diffusion method for
the emergent spectrum. The solar abundances of \citet{mcm83} were employed
in all the models, as was an incident power-law radiation field
(defined from 1~eV to 100~\kev) with photon index $\Gamma=2$. The
following ions were treated: \ion{C}{5} \ion{--}{7}, \ion{N}{6} \ion{--}{8},
\ion{O}{5} \ion{--}{9}, \ion{Mg}{9} \ion{--}{13},
\ion{S}{11} \ion{--}{15} and \ion{Fe}{16} \ion{--}{27}.

Each model started with a constant density slab ($n_{\mathrm{H}} =
10^{15}$~cm$^{-3}$) defined to have a total Thomson depth of
$\tau_{\mathrm{T}} = 6.1$. At this optical depth almost all of the
high energy photons will interact at least once with the
gas. Beginning at a depth $\tau_{\mathrm{step}}$ the density was then
multiplied by a factor $f_{\rho}$ to create an artificial drop or jump
in the density profile at that position ($\tau_{\mathrm{T}}$ remained
fixed at 6.1 by adjusting the physical width of the altered
zones). The increase or decrease in density was done discontinuously:
there are no zones with densities intermediate between
$n_{\mathrm{H}}$ and $f_{\rho}n_{\mathrm{H}}$. The width of the step
was arbitrarily defined as $\Delta \tau = k \tau_{\mathrm{step}}$,
where $k$ is a constant. This results in the steps becoming wider as
they are placed deeper which potentially allows for stronger
observable effects to the reflection spectra.

We chose to calculate reflection spectra for 4 different ionization
parameters: $\xi =$125, 250, 500, \& 1000~erg~cm~s$^{-1}$. These
values bracket the transition from a neutral 6.4~\kev\ \fe\ line to an
ionized He-like 6.7~\kev\ line \citep{ros99}. The soft X-ray features in the
reflected emission, such as \ion{O}{8}~Ly$\alpha$, also vary greatly
over this range of $\xi$ \citep*{brf02}. For each value of $\xi$, reflection
spectra were calculated for steps placed at $\tau_{\mathrm{step}} =$
0.1, 0.5, 1, 1.5 \& 2, with $f_{\rho}$ ranging from 10$^{-3}$,
10$^{-2}$, 0.1, 0.5, 2 \& 10. The width of each step was
controlled by the value of $k$, and for each $(\tau_{\mathrm{step}},
f_{\rho})$ pair, models were computed for $k =$ 0.2, 0.5, 1 \&
2. Thus, 120 different types of density steps were calculated for
each $\xi$. All but two models converged successfully, and the
resulting reflection spectra were compared to an unaltered constant
density model to assess the significance of any changes (\S
\ref{sub:toyres}).

\begin{figure}
\plottwo{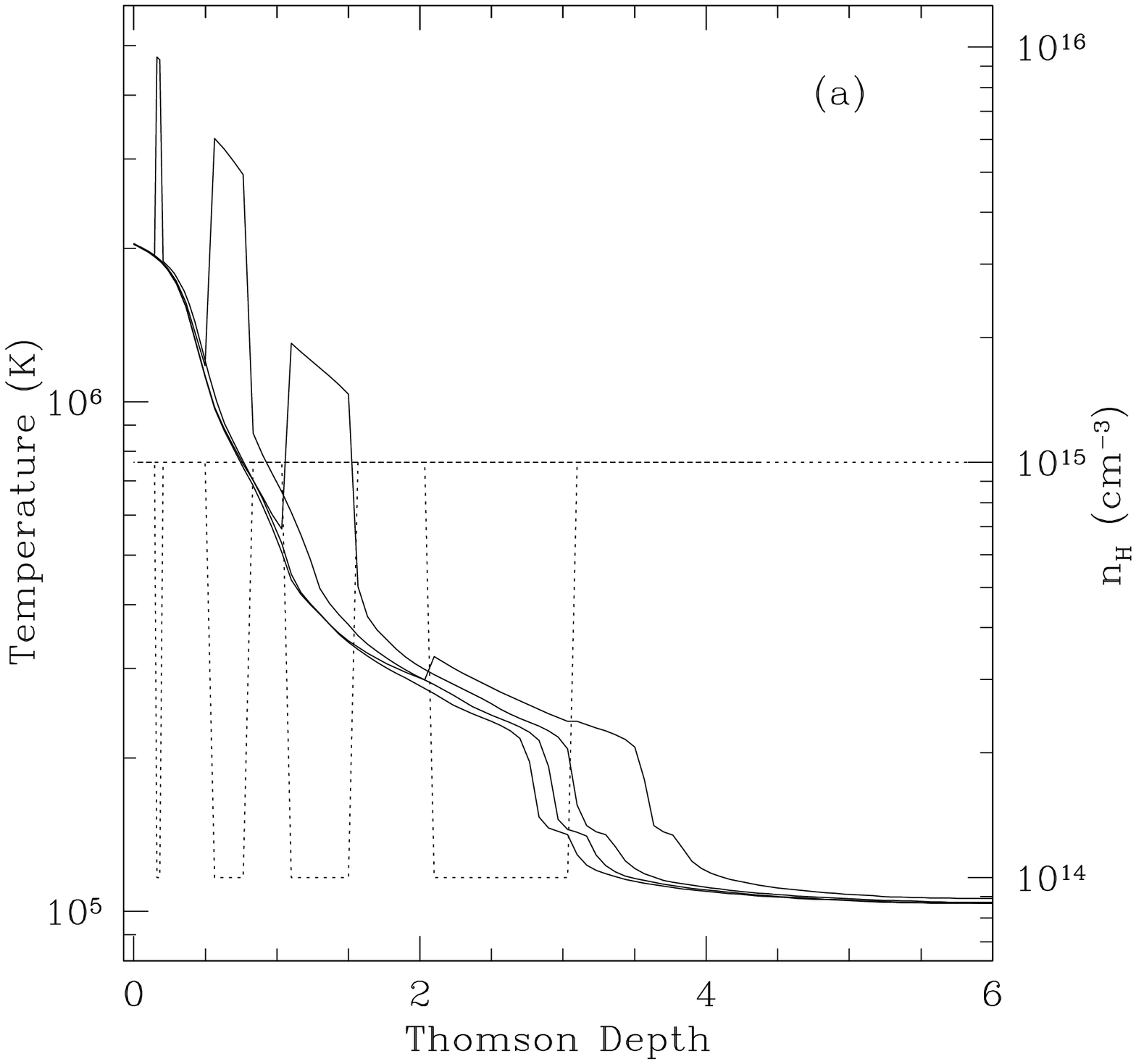}{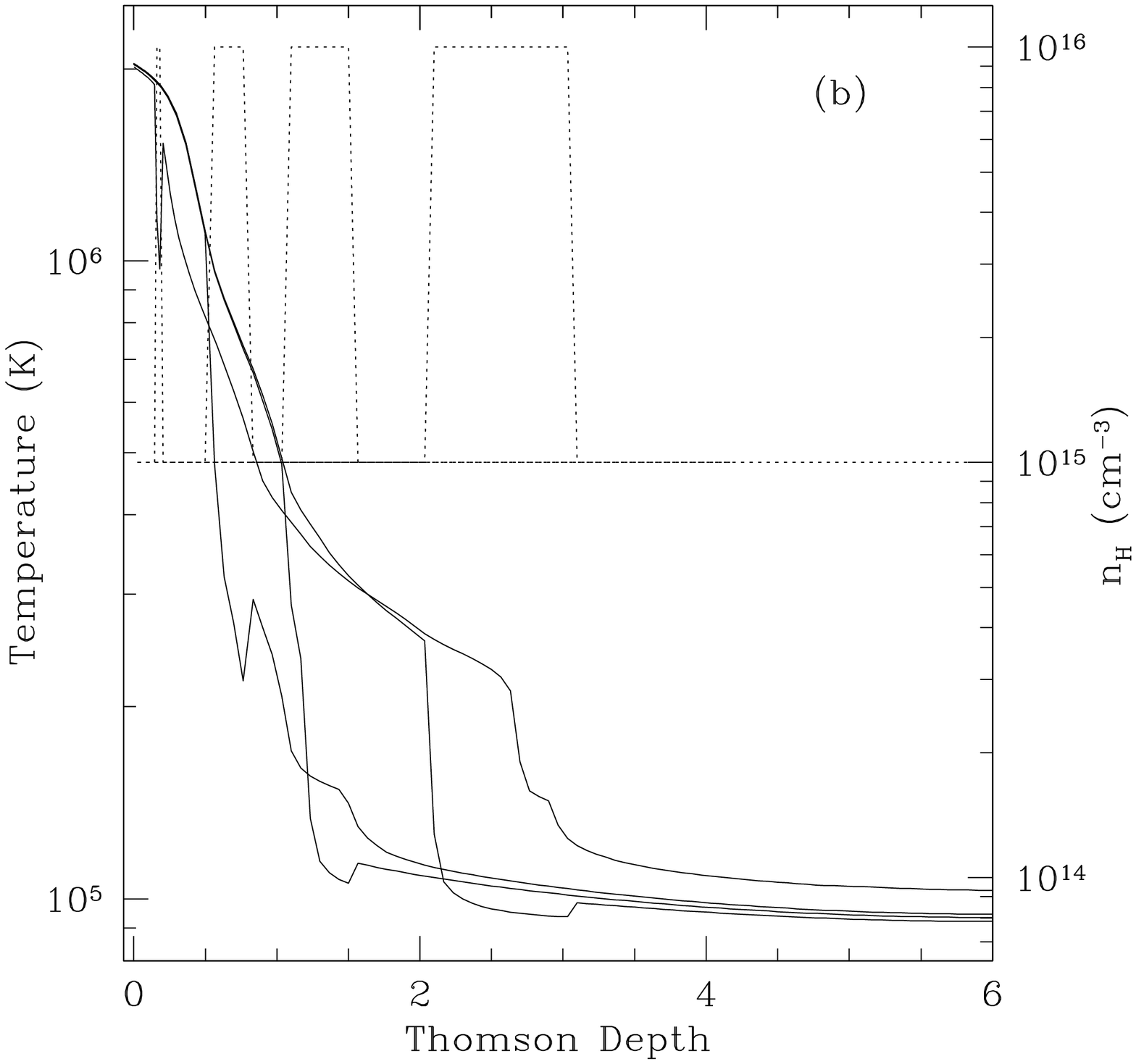}
\caption{Gas temperature (solid lines) \& number density (dotted
  lines) as a function of $\tau_{\mathrm{T}}$ for eight
  $\xi=250$~erg~cm~s$^{-1}$ models. (a) Results for the case when
  $f_{\rho}$=0.1. Steps were placed at $\tau_{\mathrm{step}}=0.1, 0.5,
  1$ \& $2$, each with $\Delta \tau=0.5 \tau_{\mathrm{step}}$. The gas
  temperature increases at the position of the drop in density, but
  otherwise there is little impact in the overall shape of the
  temperature profile. (b) As in (a), except $f_{\rho}=10$. In this
  case, the temperature falls at the jump in density, but, because
  of the increased cooling rates at lower $T$, it does not recover to a
  similar overall profile.}
\label{fig:tempanddens}
\end{figure}

As an illustration of the technique, Figure~\ref{fig:tempanddens}
plots the temperature and density profiles as a function of
$\tau_{\mathrm{T}}$ for eight $\xi=250$~erg~cm~s$^{-1}$ models. The
results are shown for $f_{\rho} = 0.1$ in
Fig.~\ref{fig:tempanddens}(a) and $f_{\rho}=10.0$ in
Fig.~\ref{fig:tempanddens}(b). In each plot the dotted lines show the
density profiles for four different steps (occurring at
$\tau_{\mathrm{step}}=0.1, 0.5, 1$ \& $2$) inserted into the slab,
while the solid lines denote the final temperature profile. All the
steps shown here have a width of $\Delta \tau = 0.5
\tau_{\mathrm{step}}$ (i.e., $k=0.5$). The gas temperature reacts as
expected: it increases by a factor of few within a density drop, and
decreases within a jump. However, the effect of an increase in
$n_{\mathrm{H}}$ seems to have a greater impact on the overall
temperature profile than a sudden drop in
density. Fig.~\ref{fig:tempanddens}(a) shows that outside of the step,
the temperature profiles are roughly similar, and indeed closely
resembles the one from a constant density model (not shown). On the
other hand, Fig.~\ref{fig:tempanddens}(b) shows that the temperature
profiles are quite different for the four models. The differences
between these two cases illustrates the major difference between an
increase or a decrease in density. If the density changes by a factor
$f_{\rho}$ then the ionization parameter in that region is changed by
$1/f_{\rho}$. When $f_{\rho} < 1$, the gas in the altered zones is
more easily ionized which raises the temperature, but, most
importantly, the recombination and free-free emissivity rates (which
scale as $n_{e}^2 \approx n_{\mathrm{H}}^2$) are reduced. It is this
increase in cooling when the gas becomes denser which has the largest
impact on the temperature profiles seen in
Fig.~\ref{fig:tempanddens}(b). However, the ionized gas in a low
density region may be a strong emitter in some energy bands (e.g., the
6.7~\kev\ \fe\ line). Furthermore, a decrease in density will also
lower the opacity to radiation emitted from below. Therefore, we now
turn to see how the impact of a sudden density change impacts the
reflection spectrum.

\subsection{Results}
\label{sub:toyres}
\begin{figure}
\centerline{
\includegraphics[angle=-90,width=0.5\textwidth]{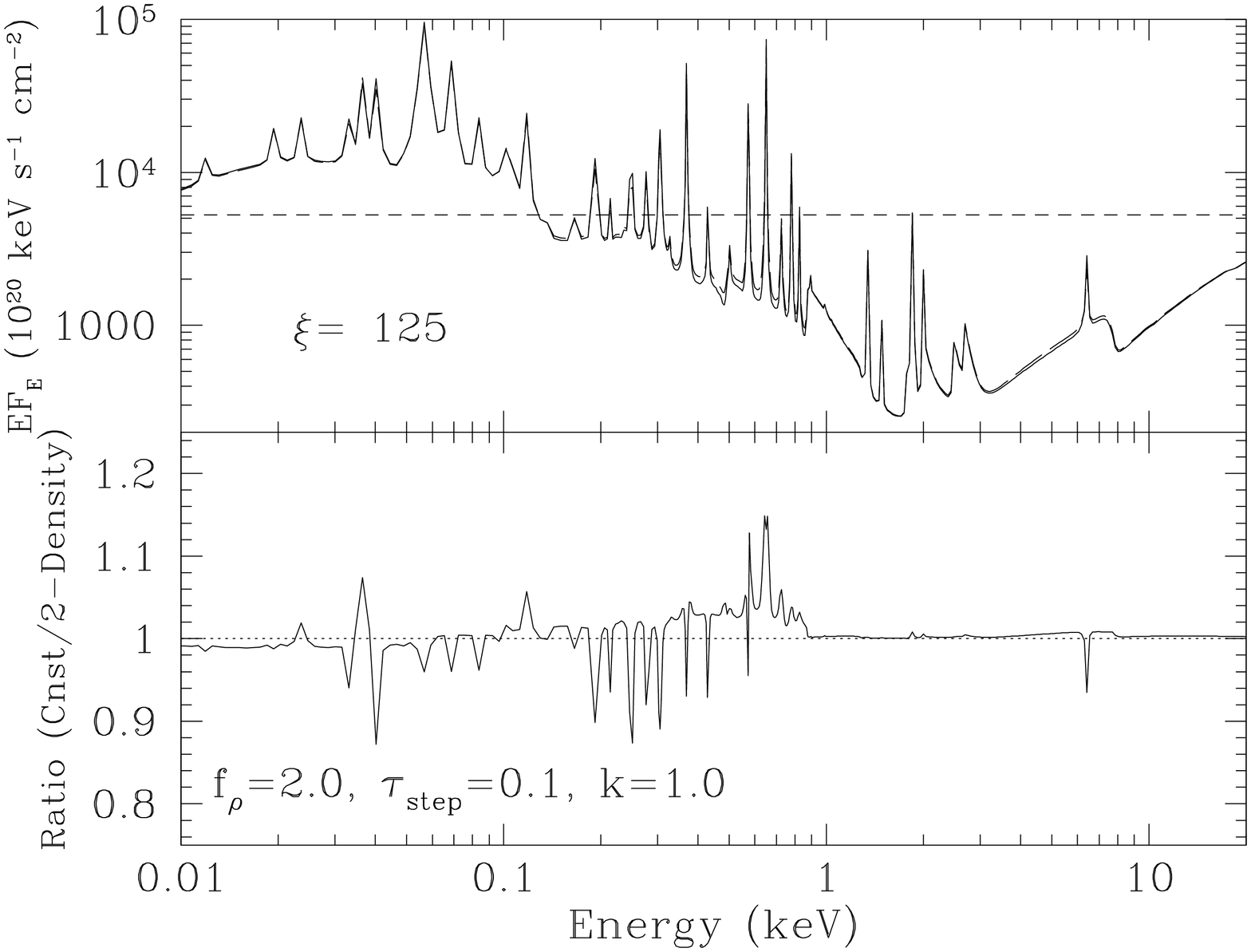}
\includegraphics[angle=-90,width=0.5\textwidth]{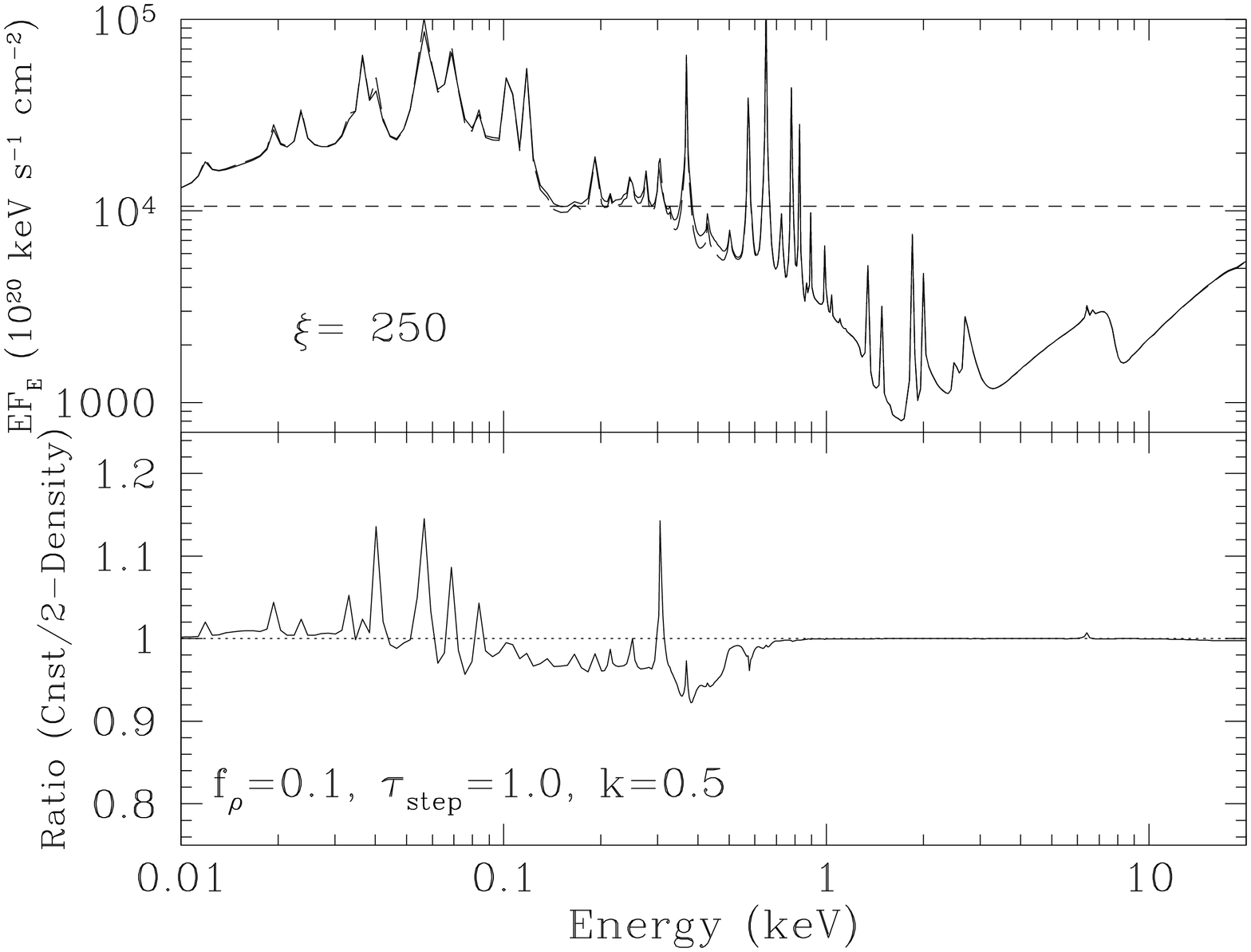}
}
\centerline{
\includegraphics[angle=-90,width=0.5\textwidth]{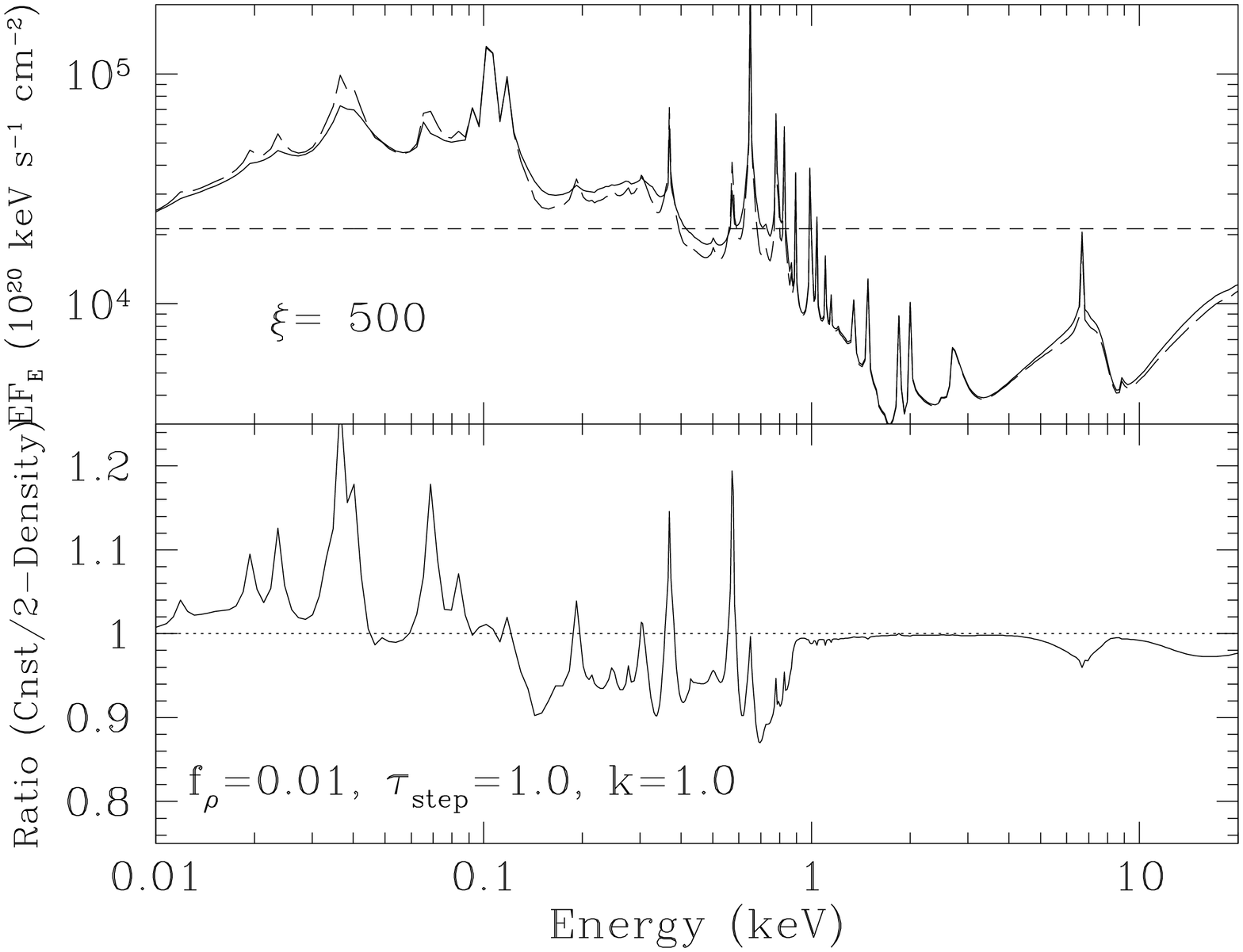}
\includegraphics[angle=-90,width=0.5\textwidth]{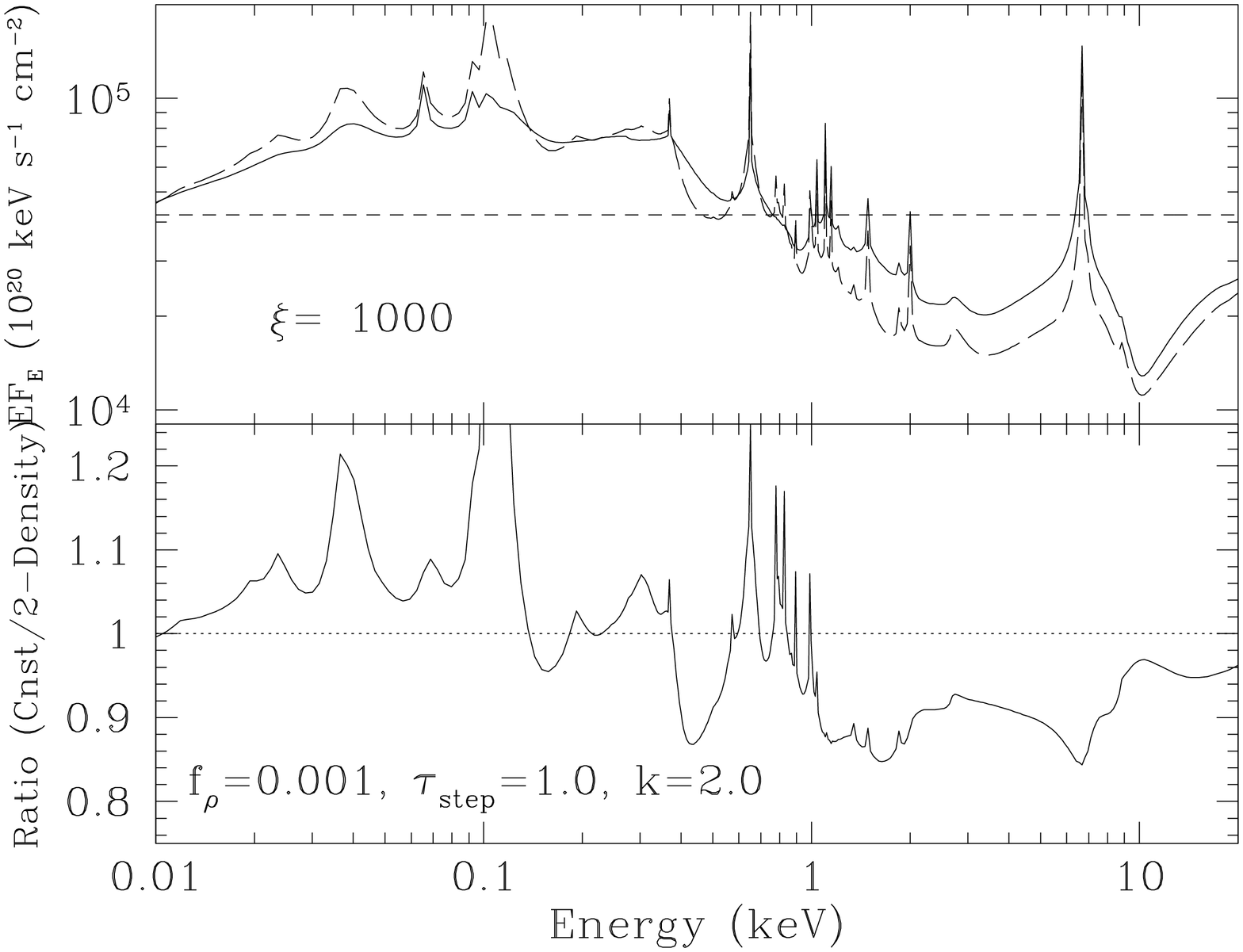}
}
\caption{Four examples of how the reflection spectra computed from the
toy models compare with those from unaltered constant density
slabs. The top panel of each plot shows the reflection spectra as the
solid line with the constant density one denoted by the long-dashed line. The
short-dashed line shows the $\Gamma=2$ incident power-law used for
both calculations. The lower panel plots the ratio of the two
reflection spectra defined as (constant density/two-density model). With AGN
the power-law is also observed along with the reflection spectrum, so
the $\Gamma=2$ spectrum was added to the reprocessed emission prior to
the ratio being calculated (i.e., they both have a reflection fraction
of unity). In this way, we are comparing the spectra as they may be observed.}
\label{fig:comparespect}
\end{figure}

As a first step in determining the effects of a step in density on the
reflection spectrum, we simply overplot the predicted emission with
that from a constant density slab under the same illumination
conditions. Four examples of this comparison, one for each value of
$\xi$, are shown in the upper panels of Figure~\ref{fig:comparespect}. Since in
some cases any differences between the two spectra are difficult to
see by eye, we plot in the lower panels the ratio between the constant
density and the inhomogeneous model spectra. Prior to the ratio being
calculated the $\Gamma=2$ power-law was added to each spectrum
so that each had a reflection fraction of unity. This was
done in order for the plot to more accurately represent the ratio of
two observed spectra (in AGN \& GBHCs the power-law continuum is
observed along with the reflected emission).

Figure~\ref{fig:comparespect}, while only showing four examples, does
demonstrate a number of useful points. First, although there are
changes, the reflection spectra
from the toy inhomogeneous models are not dramatically different from
the constant density results; that is, they still look qualitatively like the
spectra first presented by \citet{ros93}. Secondly, the three variables
regarding the density step (its placement, depth/height, and width),
as well as the value of $\xi$, all affect the deviations observed in
the spectra. For example, the step illustrated in the
$\xi=250$~erg~cm~s$^{-1}$ panel of Fig.~\ref{fig:comparespect}, which
corresponds to one of the models shown in
Fig.~\ref{fig:tempanddens}(a), results in a negligible change to the
reflection spectra above $\sim 0.7$~\kev. The only major differences
in the reflection spectrum are a decrease in intensity of low-energy
\ion{Si}{11}, \ion{O}{5}, and \ion{C}{5} lines. In this case, the drop
in energy was placed too deep into the layer for it to make much
difference to the reprocessed emission. However, at larger ionization
parameters, the incoming X-rays penetrate further into the slab, and
so a step at $\tau_{\mathrm{step}}=1$ can have a large impact on the
reflection spectrum. This is seen in the $\xi=500$~erg~cm~s$^{-1}$
panel of Fig.~\ref{fig:comparespect}, but more dramatically in the
$\xi=1000$~erg~cm~s$^{-1}$ plot. In both these cases, the low density step
alters the reflection spectra so that they have a shape that resembles
ones with a larger value of $\xi$ (less absorption, broader spectral
features). These two plots show that the reflection spectra from a
inhomogeneous slab can be a mixture of ionization parameters, similar
to ones calculated from hydrostatic atmospheres \citep{brf01}. Of course, when
the step causes an increase in density, as is shown in the
$\xi=125$~erg~cm~s$^{-1}$ panel, then a spectrum typical of a lower
ionization parameter (stronger absorption and low-energy emission
lines) is mixed in with the results. 

Although Fig.~\ref{fig:comparespect} does show that density
inhomogeneities can cause observable effects in the reflection
spectra, the results are largely qualitative and selective (there are
$\sim 475$ other possible panels). To better quantify the results, and
to obtain a more general overview of the parameter space, we
calculated the \fe\ and \oeight\ Ly$\alpha$ EWs from the total
(reflected+incident) spectrum for each two-density model and compared it to
the values from the corresponding constant density models. The EWs
were calculated by integrating the spectrum between 6 and 7~\kev\ for
\fe, and between 0.6 and 0.7~\kev\ for \oeight\
Ly$\alpha$. Figure~\ref{fig:ratiopanel} shows the results of this
exercise, where for each $\xi$ we plot the EW ratio (defined as two-density
model/constant density) for both the \fe\ (black) and \oeight\ (blue)
lines as a function of $\tau_{\mathrm{step}}$. Each plot contains 6
panels for the different values of $f_{\rho}$, and the different style
of lines within the panel denotes a particular value of $k$.  There is
plenty of information in this figure, and it is worth going through it
in detail, but before doing so there is one general point that can be
made. In almost every instance the magnitude of the EW ratio shown in
Fig.~\ref{fig:ratiopanel} is proportional to the value of $k$; that
is, the wider the step, the greater the change to the reflection
spectrum for those values of $f_{\rho}$ and $\tau_{\mathrm{step}}$.

Starting with the $\xi=125$~erg~cm~s$^{-1}$ plot, we find that because
of the relatively weak illumination, the density change generally only affects
the EWs for $\tau_{\mathrm{step}} < 0.5$. The largest changes to the \fe\
EW are when $f_{\rho}=0.1,\, 2$ or $10$. In the first case, the
increase in EW is due to the appearance of a 6.7~\kev\ component to
the \fe\ line, while the later two instances are a result of a
stronger 6.4~\kev\ component. The reason why the ionized line appears
in the $f_{\rho}=0.1$ models but not in the $f_{\rho}=0.01$ or $0.001$
cases is that the Fe is almost entirely fully stripped in the very low
density zones and so does not contribute much emission (as do the
other metals). The \oeight\ EW varies only slightly, except for when
$f_{\rho}=10$ where it drops by $\sim 30$\% due to a low ionization
fraction in the overdense region.

A similar pattern is seen in the $\xi=250$~erg~cm~s$^{-1}$ models,
with the changes now occurring for $\tau_{\mathrm{step}} < 1$. There
are significant variations in the \fe\ EW for most values of
$f_{\rho}$, with the largest changes ($>$ factor of 8) occurring in
the $f_{\rho}=0.1$ case. The reflection spectrum for the
$\xi=250$~erg~cm~s$^{-1}$ constant density model (seen as the dashed
line in Fig.~\ref{fig:comparespect}) has a very weak \fe\ line due to
Auger destruction \citep*{rfb96}. Therefore, changes to the density
which result in either a 6.7~\kev\ line ($f_{\rho}=0.1$) or a strong
6.4~\kev\ line ($f_{\rho}=10$) cause a significant increase in the
EW. At this value of $\xi$ it seems to be difficult to increase the
\oeight\ EW by density steps, but it can be lowered by nearly a factor
of two when $f_{\rho}=10$.

Increasing the illumination level so that $\xi=500$~erg~cm~s$^{-1}$
results in some changes in the reflected emission with a step down at
$\tau_{\mathrm{step}} \sim 2$, but, from Fig.~\ref{fig:ratiopanel}, we
see that the most substantial changes still occur at
$\tau_{\mathrm{step}} \la 0.5$. The \fe\ EW differs significantly from
the constant density value only when $f_{\rho}=0.1$ or $0.5$. In these
cases the iron in the low density zones becomes highly ionized, but
not fully stripped, increasing the column of the He-like species. The
line is also broader due to Compton down-scattering. Again, we see
that the spectrum takes on the properties of one from a larger
$\xi$. This is also reflected in the changes in the \oeight\ EW, which
is much smaller than the constant density model for
$f_{\rho}=0.001$--$0.5$. However, the opposite behavior (a larger
\oeight\ EW and a smaller \fe\ EW) is seen when the step causes an
increase in density and low-$\xi$ features are mixed into the
reflection spectrum. In most of those models, the 6.7~\kev\ Fe line
and the \oeight\ Ly~$\alpha$ line are reduced and enhanced in
intensity, respectively, although an interesting counter-example is
seen when $f_{\rho}=10$ and $\tau_{\mathrm{step}}=0.5$. In this model,
the overdense region is placed at a depth so that both a 6.4~\kev\ and
a 6.7~\kev\ \fe\ line is emitted, causing an increase in the
EW. Similarly, the \oeight\ EW falls because of a significant
transition to \ion{O}{7} emission.

Finally, the most highly ionized runs ($\xi=1000$~erg~cm~s$^{-1}$)
show little variation in the \fe\ EW when $f_{\rho} < 1$. The only
exceptions are when the density deficit is near the surface and the
high ionization parameter in these zones decreases the amount of
6.7~\kev\ emitting ions. As a result, the line EW falls by about
25\%. On the other hand, when $f_{\rho}=10$, the \fe\ EW can drop by
over a factor of two because the overdense region stops the ionization
front at a much shallower depth in the slab. The \oeight\ EW is always
smaller than its constant density value when $f_{\rho} < 1$, but can
increase by over a factor of 2 (up to 110~eV) when there is a density
enhancement just beneath the surface.

In summary, we find that toy models which place a top-hat function of
variable width and depth into a constant density slab, can produce
observable effects on the resulting reflection spectra. Generally the largest
impact on the spectrum, as measured by the \fe\ and \oeight\ EWs,
occurs when the step is placed at $\tau_{\mathrm{step}} \sim 0.5$ into
the layer. Furthermore, changing the density by only relatively small
factors of 2--10 can produce significant effects. The resulting
reflection spectra are reminiscent of the one produced by hydrostatic
models in so far as they show features from a mixture of ionization
parameters. 

\subsection{Fits with Constant Density Reflection Spectra}
\label{sub:toycomp}
\begin{figure}
\centerline{
\plottwo{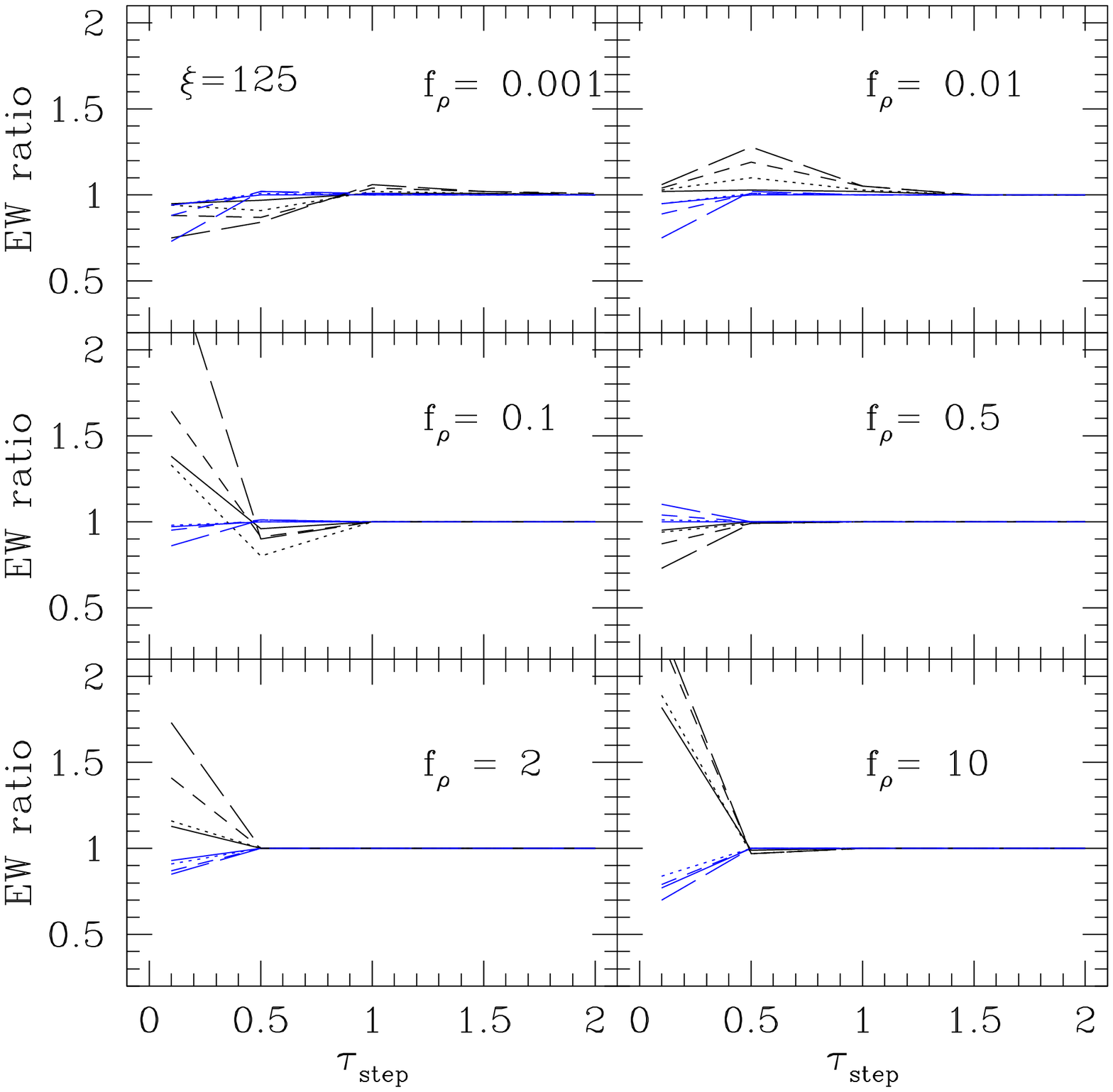}{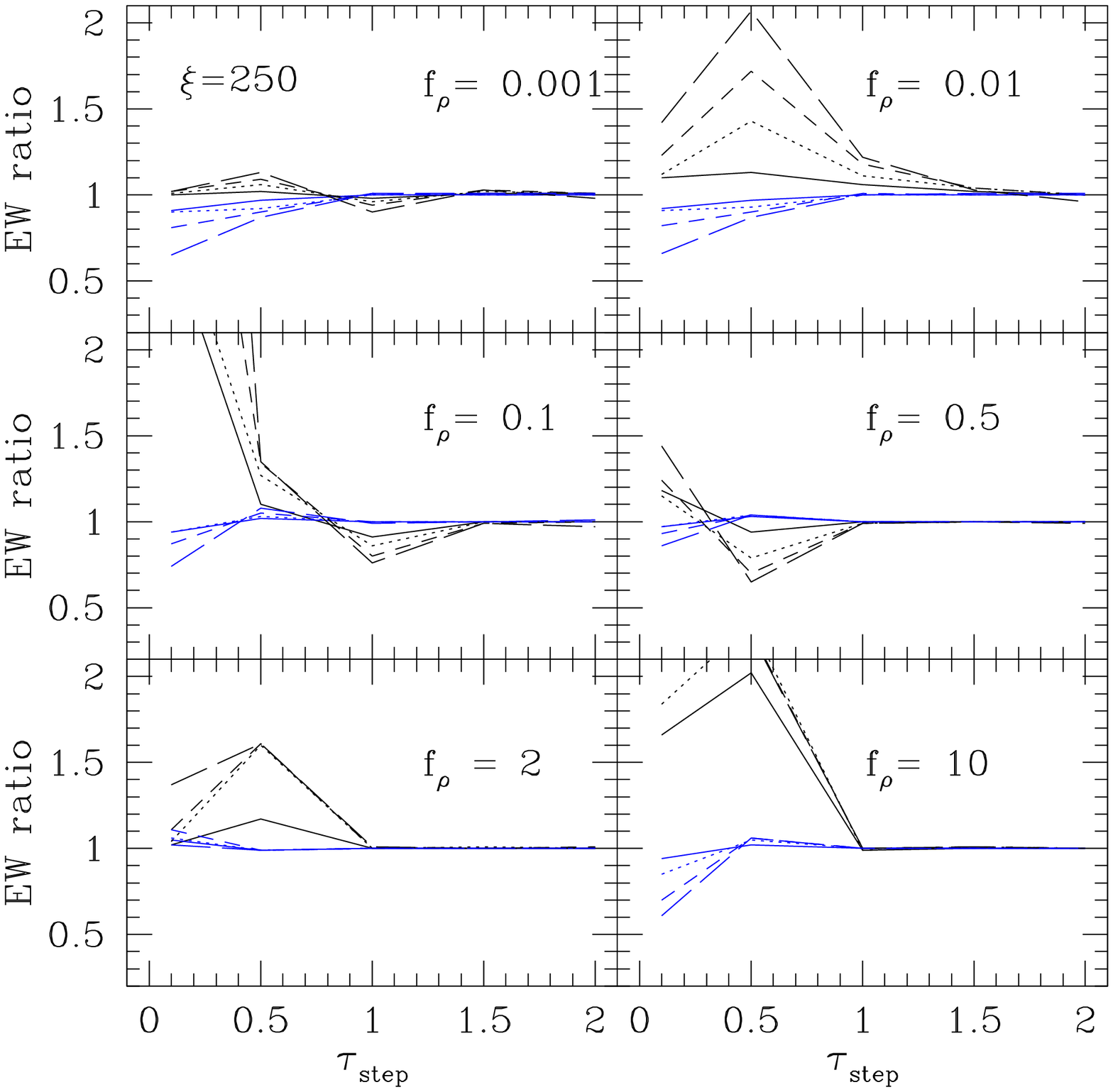}
}
\centerline{
\plottwo{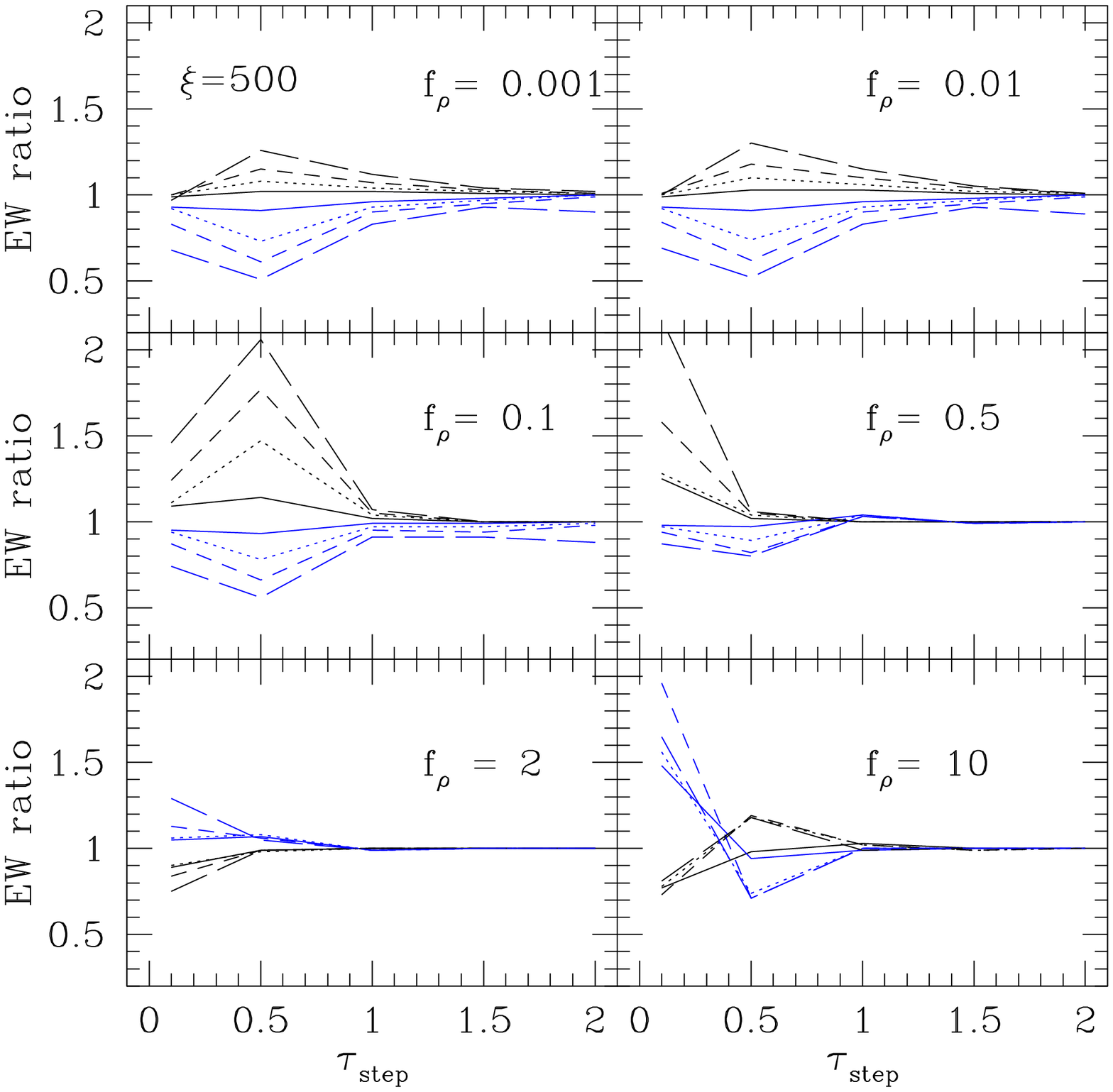}{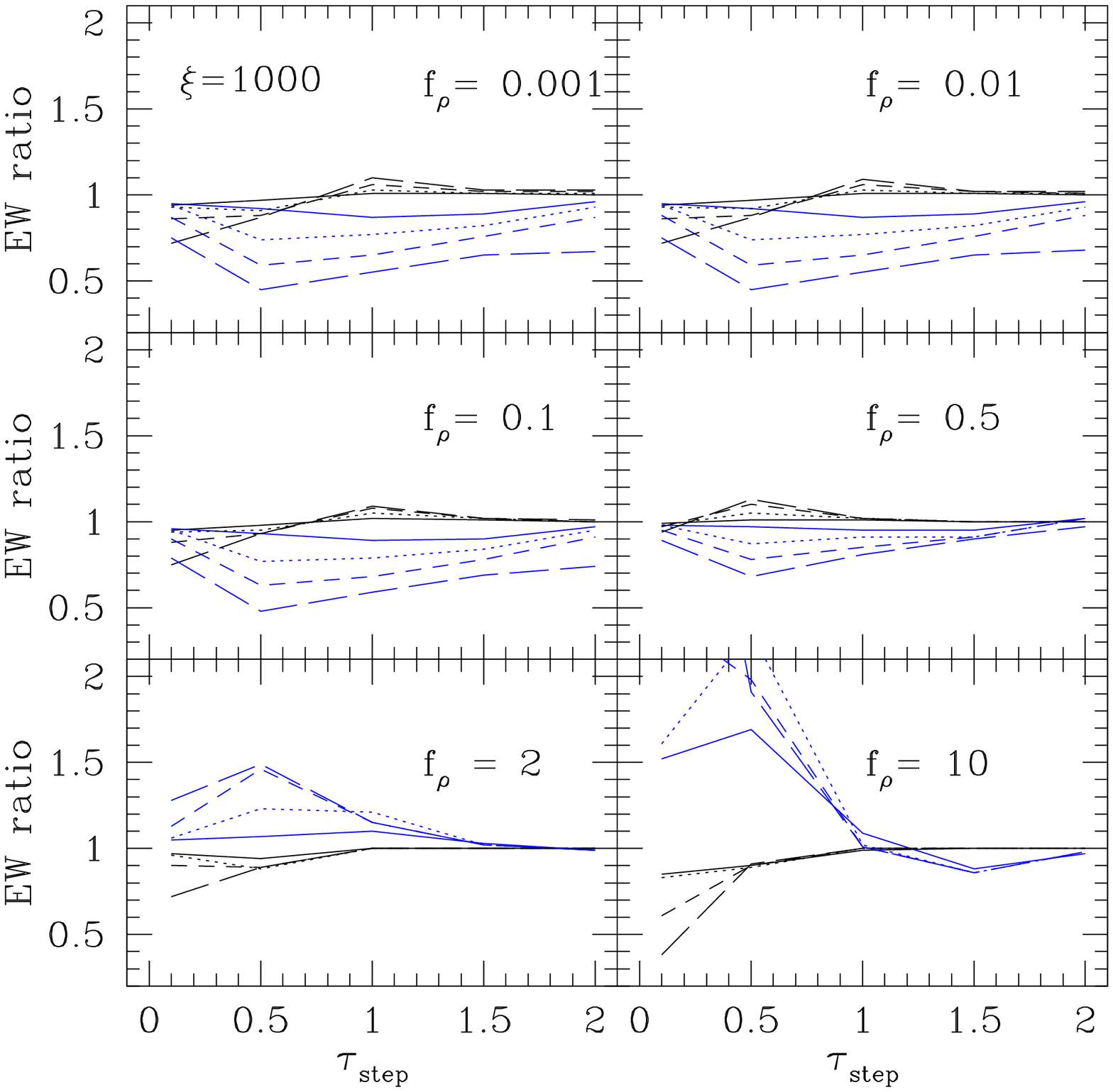}
}
\caption{These plots show as a function of $\tau_{\mathrm{step}}$ the
  ratio of the \fe\ (black) and O~\textsc{viii} (blue) Ly$\alpha$ equivelant
  widths from the toy inhomogeneous models to those from the constant
  density models. The six panels in each plot descibe the results for
  each value of $f_{\rho}$, while the linestyles denote the different
  values of $k$: solid ($k=0.2$), dotted ($k=0.5$), short-dashed
  ($k=1$) and long-dashed ($k=2$). The EWs were calculated from the
  total (reflection+incident) spectra. See the text for the discussion.}
\label{fig:ratiopanel}
\end{figure}

The EW ratios shown in Fig.~\ref{fig:ratiopanel} indicate that density
inhomogeneities in the surfaces of accretion disks may have an
observable effect on their reflection spectra. Those measurements,
however, only use a fraction of the information contained in the model
spectra. In this section, we take several example reflection spectra
calculated from the two-density models and fit them with ones from
the constant density models. This method will allow us to quantify any
differences between the two scenarios by using all the information
over a given energy range. The exercise may also be indicative of how
real inhomogeneities will manifest themselves when fitting data with
simple models.

\begin{figure}
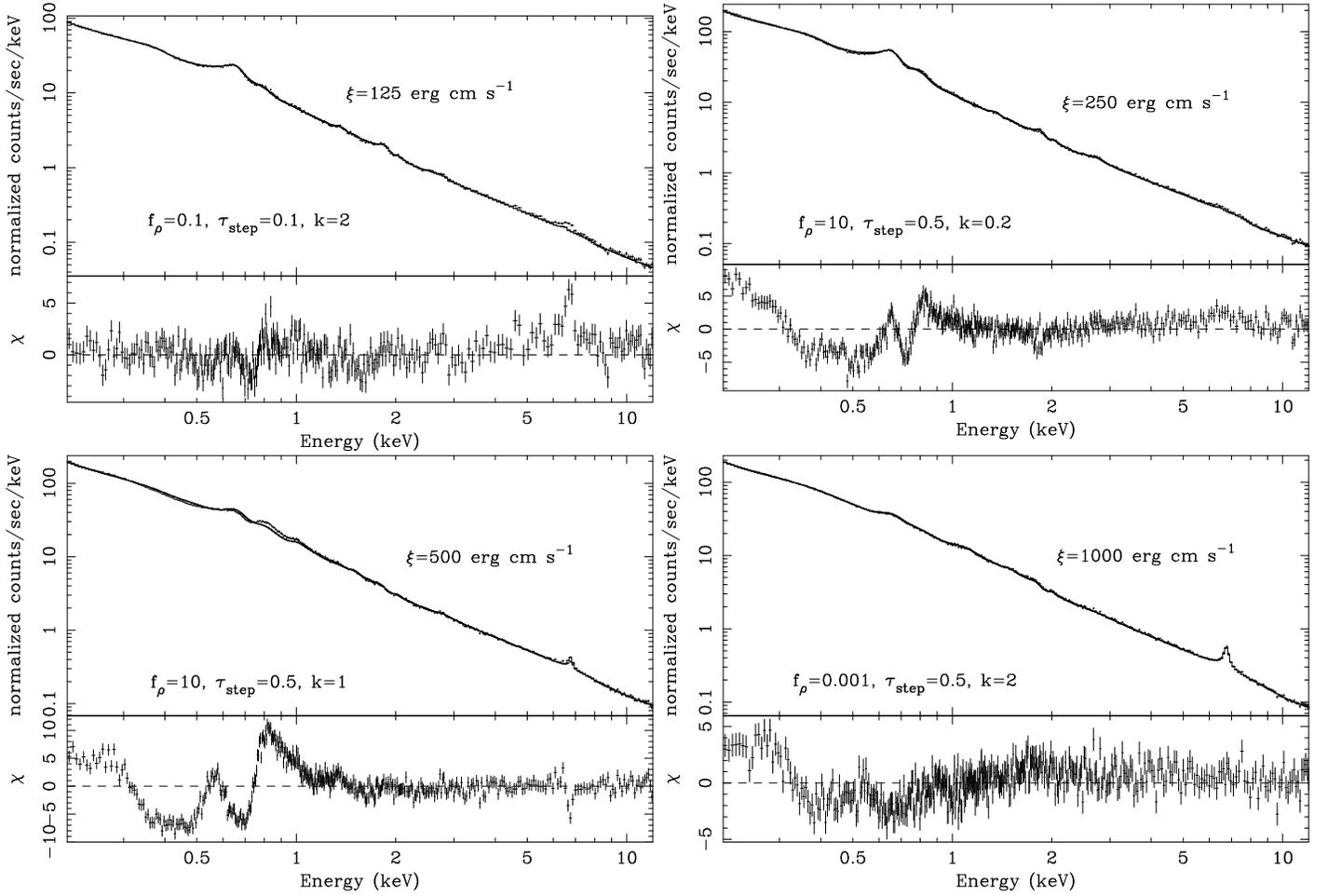

\centerline{
\includegraphics[angle=-90,width=0.5\textwidth]{f4a.eps}
\includegraphics[angle=-90,width=0.5\textwidth]{f4b.eps}
}
\centerline{
\includegraphics[angle=-90,width=0.5\textwidth]{f4c.eps}
\includegraphics[angle=-90,width=0.5\textwidth]{f4d.eps}
}
\caption{Simulated count spectra and residuals (in units of standard
  deviations) to fits of four two-density
  models with constant density reflection spectra. The fit
  parameters are given in Table~\ref{table:stepfits}.}
\label{fig:stepfits}
\end{figure}

We chose models that showed large changes in its \fe\ or \oeight\ EW
from each of the four ionization parameters considered (the exact
model parameters are listed in the note to
Table~\ref{table:stepfits}). Simulated observations were made with
these model spectra by using the 'fakeit' command in
\textsc{xspec}v.11.2.0bp \citep{arn96}. The canned \textit{XMM-Newton}
EPIC-pn response matrix \texttt{epn\_sw20\_sdY9.rmf} was used to
generate the data assuming an exposure time of 40~ks. As with the EW
calculation in the previous section, the incident power-law was added
to the reflection spectra so that each simulated spectrum had a
reflection fraction, $R$, of unity. Each faked dataset was then fit
with a grid of constant density models (calculated with the same
version of the reflection code) between 0.2 and 12~\kev\ as well as
between 2 and 10~\kev. The fit parameters were the photon index
$\Gamma$, the ionization parameter $\xi$, the reflection fraction $R$,
and an arbitrary normalization. The results are listed in
Table~\ref{table:stepfits}, and the residuals are shown in
Figure~\ref{fig:stepfits}.

This particular $\xi=125$~erg~cm~s$^{-1}$ model was chosen because the
\fe\ EW increased by over a factor of 2 due to the appearance of a
6.7~\kev\ line. In the 0.2--12~\kev\ energy band, the constant density
model does a decent job of recovering the ionization parameter, but the
value of $R$ is too low. In contrast, the best fit in the 2--10~\kev\
range has a much larger $\xi$ (466~erg~cm~s$^{-1}$) and
$R$. Evidently, the spectrum in this energy band mimics one from a more
ionized reflector. The residuals to the 0.2--12~\kev\ fit are shown in
Fig.~\ref{fig:stepfits} and shows that the low value of $R$ is likely
due to the overprediction of the \oeight\ line (as seen in
Fig.~\ref{fig:ratiopanel}). Moreover, the ionized \fe\ line cannot be
accounted for by this low $\xi$. Ignoring the low energy data
increases the reflection fraction as the spectrum attempts to account
for the enhanced Fe emission. In both cases, the continuum is fit
adequately, but the differences in line emission produce significant
residuals.

Statistically, the fit to the $\xi=250$~erg~cm~s$^{-1}$ two-density model
over the 0.2--12~\kev\ band is much worse than for the previous
case. The best fit ionization parameter is lower than the true one,
and so overpredicts the strength of the soft emission lines between
0.3 and 0.6~\kev\ (Fig.~\ref{fig:stepfits}). According to
Fig.~\ref{fig:ratiopanel}, the \fe\ EW in this two-density model was $\sim$2
times greater than the corresponding constant density model (although
at this $\xi$ the EW is only 20~eV), but that the change in the \oeight\ EW
was marginal. Thus, in this case we find that the overdense region
lowers the effective ionization parameter of the reflection spectrum,
but this is only discernible through a fit to the soft X-ray emission. Indeed,
we find a very good and statistically acceptable fit to this two-density
model over the 2--10~\kev\ band. The best fit $\xi$ and $R$ are only
slightly larger than their true values to account for the enhanced
\fe\ line. 

The two-density model chosen from the $\xi=500$~erg~cm~s$^{-1}$ suite of
calculations is another overdense model with the step appearing at
$\tau_{\mathrm{step}}=0.5$ below the surface. As discussed in
\S~\ref{sub:toyres}, this model includes both 6.4 and 6.7~\kev\ iron
emission lines, as well as a weaker-than-expected \oeight\
line. Interestingly, this model gives the worst result when fit by the
constant density models, with the reduced $\chi^2 > 3$. As in the
$\xi=250$~erg~cm~s$^{-1}$ model, the poor fit is caused almost
entirely by an overprediction of the soft X-ray spectral features. The
majority of the 0.2--12~\kev\ continuum is well fit by an ionized
reflector (the best fit $\xi$ is $\sim$630~erg~cm~s$^{-1}$), but the
emission lines cannot be accounted for by this model, resulting in the low
value of $R$. The overdense region sitting just below the surface of
the reflector produces features that cannot be fit by a single-$\xi$
model. The constant density models do a much better job in the
2--10~\kev\ band, where the best fit ionization parameter is nearly
exactly the true one. However, the reflection fraction must be lowered
to fit the weaker 6.7~\kev\ Fe line. There are still some residuals in
this fit between 7 and 8~\kev\ where the overdense region in the two-density
model causes additional absorption than what is predicted by the
constant density model.

Our final example considers the case where higher-$\xi$ features are
mixed into the reflection spectrum. The $\xi=1000$~erg~cm~s$^{-1}$
model used for fitting had a step with a density drop of
$f_{\rho}=0.001$ at $\tau_{\mathrm{step}}=0.5$. The lower density
material became highly ionized and weakened both the \fe\ and \oeight\
lines (Fig.~\ref{fig:ratiopanel}). The two constant density fits both show
the effects of this dilution, as the best fit $\xi$ is greater than
the true value and $R<1$. The residuals shown in
Fig.~\ref{fig:stepfits} show that an overprediction of soft X-ray
features is again the cause for the poor fit. There is also a
significant difference in the curvature of the two spectra noticeable
between 1 and 3~\kev. This is caused by the lower absorption in the
two-density model which flattens the continuum.

\section{Photon Bubbles and Reflection}
\label{sect:bubbles}
The idealised two-density models discussed in the previous section
     show that inhomogeneities in accretion disk atmospheres may
     affect the reflection spectrum.  We next examine reflection from
     density structures arising in hydrodynamical disk models.
     Effects of the ionization processes on the density structure are
     neglected. Nevertheless, the density arrangement is completely
     novel in terms of reflection calculations, as it is the first to
     be considered which did not arise from an ad hoc \emph{a priori}
     assumption (such as constant density or hydrostatic balance).

The two-density results indicate that the reflection spectrum is
     affected by inhomogeneities within the outermost few Thomson
     depths of the disk atmosphere.  By contrast, fluctuations
     resulting from the MRI may be strongest on scales comparable to
     the disk thickness \citep*{hgb95,stn96}.  The effects of photon
     bubbles are therefore considered instead. The PBI grows most
     rapidly at short wavelengths \citep{bs01}, and in Shakura-Sunyaev
     models typically grows fastest in the disk surface layers
     (N.J. Turner \etal\ 2004, in preparation).  Exploring X-ray
     reprocessing from photon bubbles is likely to be a good first
     step toward understanding the reflection properties of
     radiation-dominated disks.

The photon bubble instability was first discussed by \citet{ar92} in
     the context of accreting magnetized neutron stars, and was shown
     by \citet{gam98} to occur in the inner, radiation-supported
     regions of magnetized accretion disks.  For magnetic pressures
     exceeding the gas pressure, and perturbation wavelengths shorter
     than the gas scale height, the instability operates when
     disturbances in radiative flux displace gas along field lines,
     leading to propagating density variations which grow over time.
     Linear growth is fastest at wavelengths shorter than the gas
     pressure scale height, but is absent at wavelengths with optical
     depths much less than unity, as these support no flux variations.
     An approximate criterion for instability is that photon diffusion
     carry a greater energy flux than radiation advected at the gas
     sound speed \citep{bs03}.

\begin{figure}
\plottwo{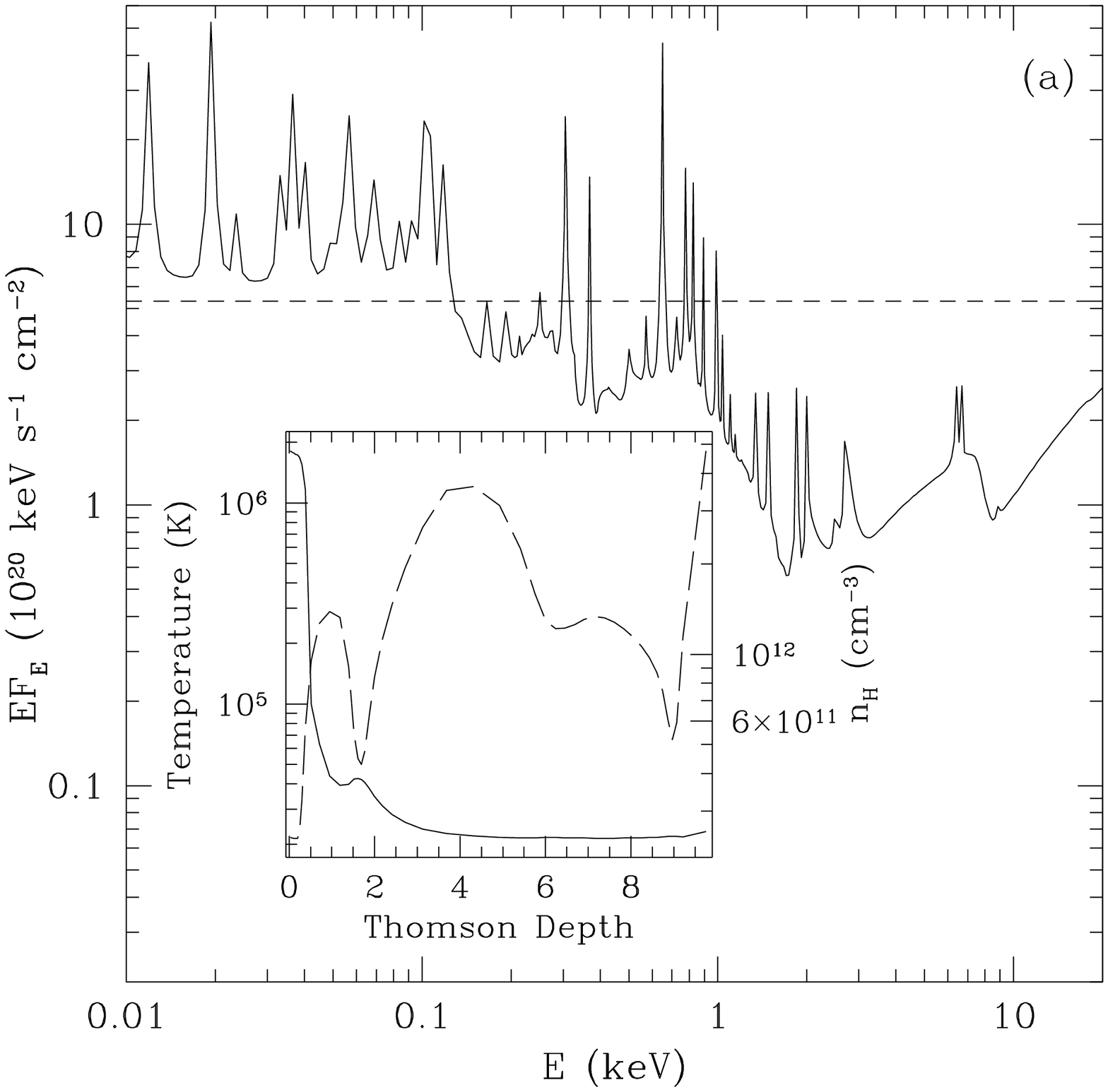}{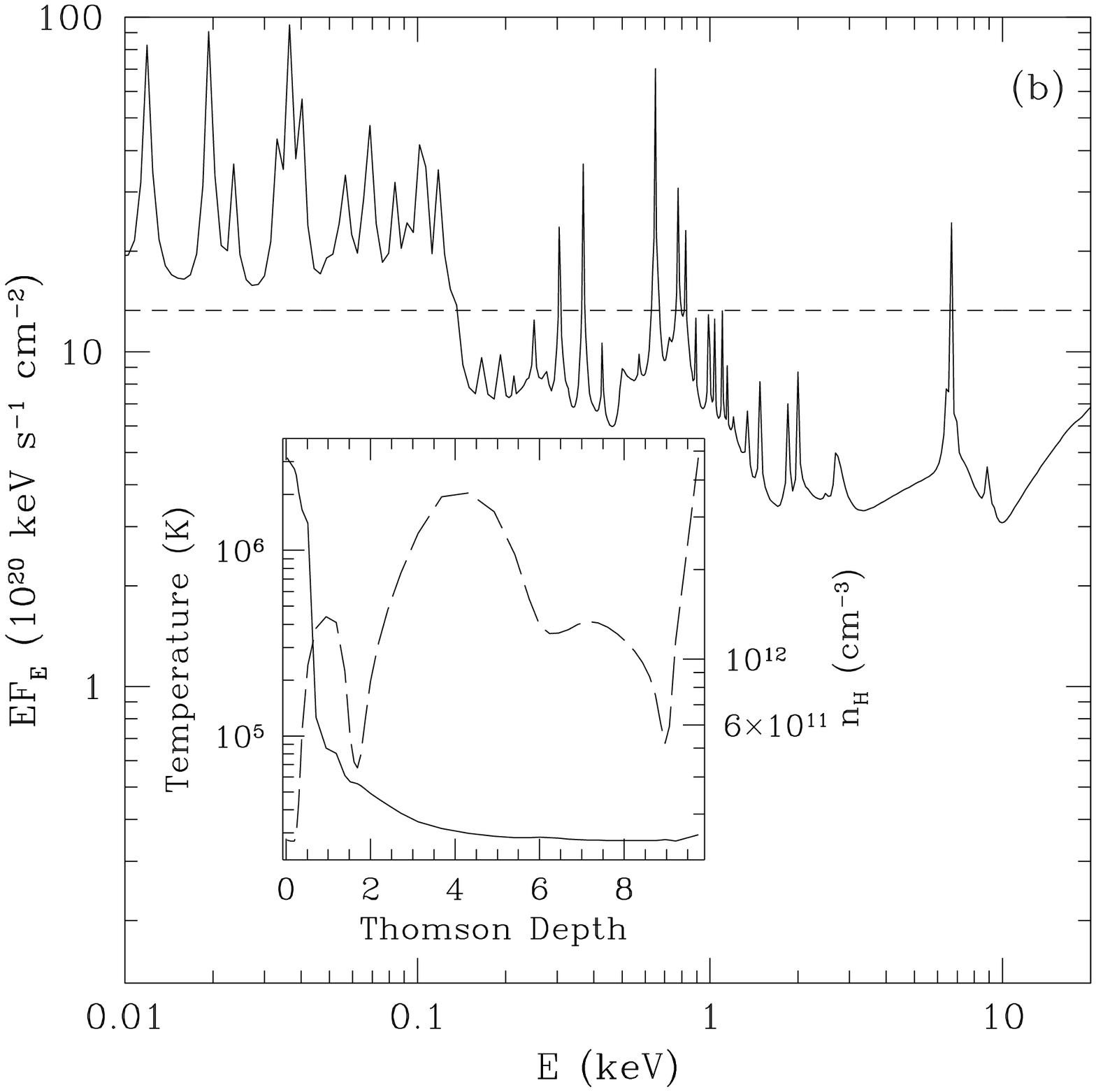}
\caption{X-ray reflection spectra (solid line) calculated from the
outer 10 Thomson depths of a photon bubble simulation. The short-dashed
line denotes the $\Gamma=2$ power-law that was incident on the
material. The insert shows the gas temperature (solid line) and number
density (dashed line) for each model. Of course, the density is the
same for each calculation. The different plots show the results for
various incident X-ray fluxes: (a)
$F_{\mathrm{X}}=10^{13}$~erg~cm$^{-2}$~s$^{-1}$, (b)
$F_{\mathrm{X}}=2.5\times 10^{13}$~erg~cm$^{-2}$~s$^{-1}$.}
\label{fig:bubblespect}
\end{figure}

Reflection spectra were calculated using a density profile taken from a
     two-dimensional radiation-MHD simulation of the growth of photon
     bubbles in a Shakura-Sunyaev model disk.  The simulation was
     carried out with the Zeus MHD code \citep{sn92a,sn92b} and its
     flux-limited radiation diffusion module \citep{ts01}.  Opacities
     due to electron scattering and bremsstrahlung were included.  The
     domain was a small patch of the disk, centered
     34~$R_{\mathrm{S}}$ from a black hole of $10^8 M_\odot$.  The
     patch extended 1.08 Shakura-Sunyaev scale-heights either side of
     the midplane, and its width was one-quarter of its height.  The
     grid consisted of $64\times 256$ zones.  The initial density and
     temperature profiles were those of a Shakura-Sunyaev model with
     $\alpha=0.06$, and accretion rate 12\% of the Eddington rate for
     a 10\% luminous efficiency.  During the simulation, no
     $\alpha$-viscosity was applied.  Differential rotation was
     neglected, so that the MRI was absent, and the material was
     allowed to cool by radiative losses.  Photons initially diffused
     from midplane to boundary in nine orbits.  The calculation was
     started with a magnetic field having a pressure 10\% of the
     midplane radiation pressure, and inclined at $45^\circ$.  After
     3.4 orbits, the PBI had developed into trains of propagating
     shocks, with density contrasts of up to two orders of magnitude.
     The vertical density profile used in the subsequent reflection
     calculations was taken from an arbitrary vertical ray, and is
     shown in the insets to Figure~\ref{fig:bubblespect}.

\subsection{Results}
\label{sub:bubres} 
As with the two-density models, this new density structure was then
irradiated by a power-law ($\Gamma=2$) of X-rays. In this case, there
is no unique ionization parameter so instead the models are
distinguished by the incident X-ray flux: $F_{\mathrm{X}}=10^{13}$,
$2.5\times 10^{13}$, $5\times 10^{13}$, and
$10^{14}$~erg~cm$^{-2}$~s$^{-1}$. The Shakura-Sunyaev flux for the assumed
disk parameters is $3 \times 10^{13}$~erg~cm$^{-2}$~s$^{-1}$. All four
models converged, and the resulting reflection spectra are presented
in Figure~\ref{fig:bubblespect}.  The gas temperature and number
density are plotted in the insets to each panel, and illustrates how
the temperature structure for each case was affected by the density
profile.

The latter three spectra all exhibit strong 6.7~\kev\ emission lines
from He-like iron, while, unusually, the most weakly illuminated model
produces both a 6.4 and a 6.7~\kev\ line of about equal strength. The
rapid drop in density below one Thomson depth causes the surface
layers to be easily ionized and able to produce a 6.7~\kev\ emission
line even with a relatively small incident flux. On the other hand,
the overdense region at $\tau_{\mathrm{T}} \approx 1 $ produces a lot
of soft X-ray line emission, even for highly illuminated
situations. Thus, the emission below $2$~\kev\ is consistently
important, despite the incident flux varying by an order of magnitude.

\subsection{Fits with Constant Density Reflection Spectra}
\label{sub:bubcomp}
\begin{figure}
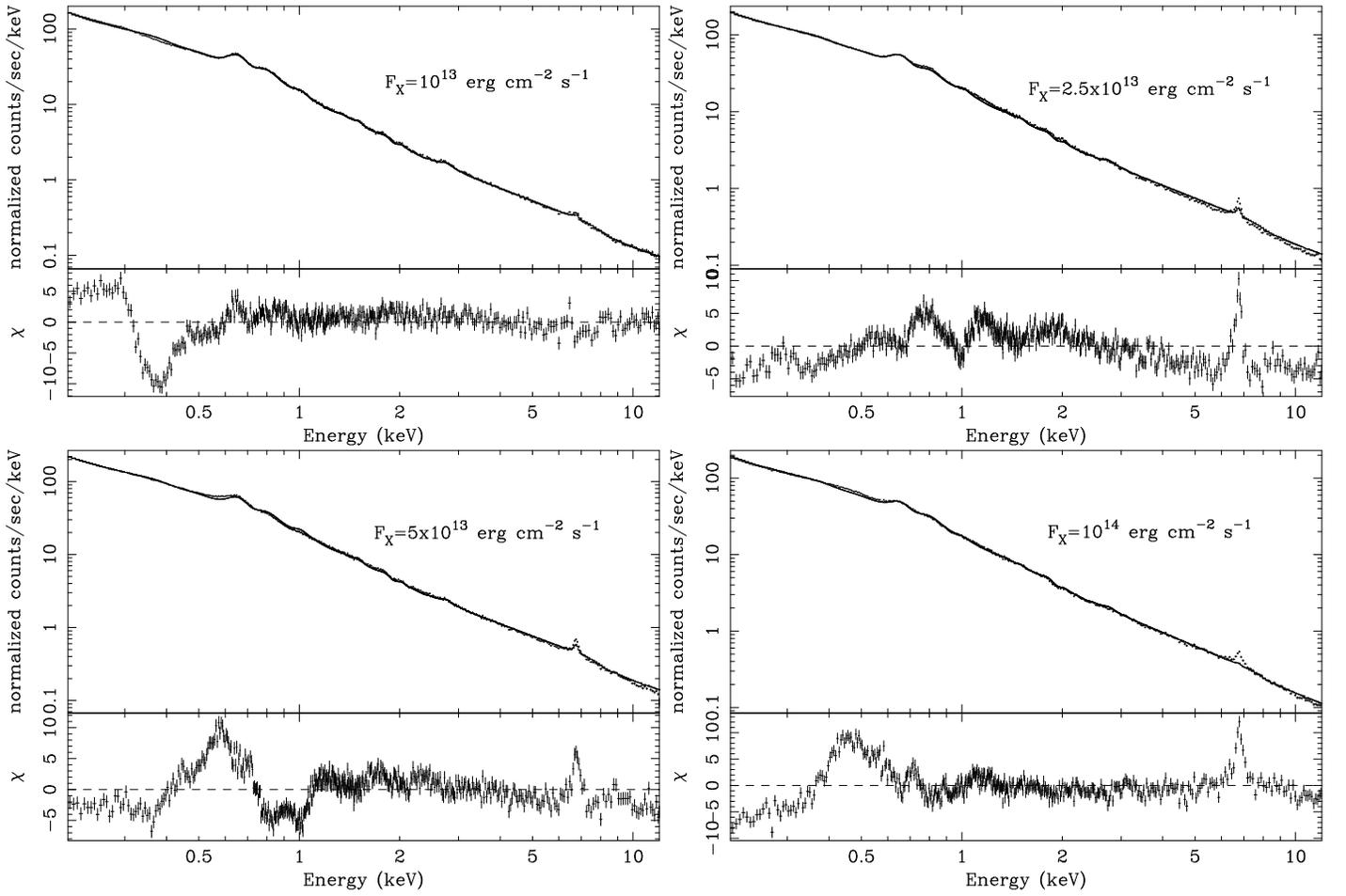

\centerline{
\includegraphics[angle=-90,width=0.5\textwidth]{f6a.eps}
\includegraphics[angle=-90,width=0.5\textwidth]{f6b.eps}
}
\centerline{
\includegraphics[angle=-90,width=0.5\textwidth]{f6c.eps}
\includegraphics[angle=-90,width=0.5\textwidth]{f6d.eps}
}
\caption{Simulated count spectra and residuals (in units of standard
  deviations) of models computed using the density cut from the photon
  bubble simulation, and then fit with constant density reflection
  spectra. The fit parameters are given in Table~\ref{table:bubblefits}.}
\label{fig:bubblefits}
\end{figure}

To quantify any effects the photon bubbles may have on the reflection
spectra we fit the models with a grid of constant density spectra. The
procedure was exactly the same as with the two-density models in
\S~\ref{sub:toycomp}: the power-law was added to the photon bubble
spectra and then the sum was used as the basis for a 40~ks
\textit{XMM-Newton} simulation. As before, constant density fits were
performed in both the 0.2--12~\kev\ and 2--10~\kev\ energy bands. A
constant density grid with $n_{\mathrm{H}}=1.4\times 10^{12}$~cm$^{-3}$
was used for fitting the photon bubble reflection spectra in order to
offset any changes in the reflection spectra due solely to the lower
density. The best fit parameters for each value of $F_{\mathrm{X}}$ are
listed in Table~\ref{table:bubblefits}, and the residuals are shown in
Figure~\ref{fig:bubblefits}.

The constant density models had difficulty fitting the spectra over
the wide 0.2--12~\kev\ energy range, with most of the trouble arising
from the soft X-ray lines. The best $\chi^2$ was found with the
$F_{\mathrm{X}}=10^{13}$~erg~cm$^{-2}$~s$^{-1}$ model, where one
ionization parameter seemed to provide a good fit to the spectrum at
energies $\ga 0.7$~\kev. The best fit $\xi$ was indicative of an
ionized slab, and, indeed, the model could not account for the small
6.4~\kev\ line predicted by the spectrum. Despite this problem, the
fit was able to recover the correct values of both $\Gamma$ and
$R$. The fit had difficulty below 0.5~\kev, however, where it
overpredicted the emission. This is most likely a result of extra
absorption (over that predicted for the fitted $\xi$) due to the
density enhancement just beneath the surface in the photon bubble
model.

Fits to the remaining three models in this energy band all resulted in
reduced $\chi^2 > 2$. A single ionization parameter could not
simultaneously account for the shape of the continuum, and the strong
soft X-ray and 6.7~\kev\ \fe\ lines. The best fit values of $R$ were
always less than the `true' value of unity. In these cases the photon
bubble models predicted significant emission from a variety of
ionization parameters, so that the resulting spectrum exhibited a
mixture of features. As a result, a single ionization parameter fit
was not adequate.

The fits to the photon bubble models improved markedly in the 2--10~\kev\ energy band
with $\xi$ increasing substantially in order to fit the \fe\
line. As a result, all four fits were statistically acceptable. Except
for the $F_{\mathrm{X}}=10^{13}$~erg~cm$^{-2}$~s$^{-1}$ model, the
reflection fractions again are still underestimated by the constant
density fits with the best fit value of $R$ decreasing with the
$F_{\mathrm{X}}$ of the model. A greater than expected amount of
absorption is the likely explanation for this effect. The density
enhancement at $\tau_{\mathrm{T}} \sim 1$ in the photon bubble model
causes more continuum absorption from oxygen then what the reflection
models expect (given, e.g., the highly ionized \fe\ line). This
enhancement becomes more important as $F_{\mathrm{X}}$ is increased
and spectral features are formed both above and below it. The extra
absorption decreases the magnitude of the continuum and emission, and thus
results in a lower $R$ when fit with the constant density
models. Aside from the low value of $R$, reflection from a uniform
slab does a good job describing the photon bubble spectra between 2
and 10~\kev.
 
\section{Discussion}
\label{sect:discuss}
Currently, the interpretation of the reflection continuum observed in
the X-ray spectra of AGN is the primary means to glean information
regarding the accretion geometry and its radiative environment. It is
therefore vital that the models of reflection, on which such
conclusions are based, explore a number of different physical
situations which may be relevant to accretion physics. Motivated by
the inhomogeneous nature of recent numerical simulations of
radiation-dominated accretion disks, we have begun a study of
examining the consequences of such density changes to the X-ray
reflection spectrum.

Our first results, presented in the previous sections, considered
reprocessing from a two-density model, where density steps or jumps were
inserted into a uniform slab, and a slice from a PBI
simulation. The new reflection spectra were then compared with ones
calculated assuming a constant density atmosphere, as these are most
frequently used in data analysis. We found that the density
inhomogeneities can result in observationally important differences
between the two cases. As in models with a hydrostatic atmosphere, the
reflection spectra can no longer be described by a single ionization
parameter, but exhibit features from a mixture of ionization
states. However, while in the hydrostatic case the spectra can be
adequately described as a diluted constant density model
because of a diffuse hot scattering layer on the surface \citep{brf01}, the
inhomogeneous models can show emission from both low and high $\xi$
material simultaneously, depending on the nature of the density
change. This gives the spectra a complexity and richness that the
hydrostatic models lacked.

The magnitude and characterization of the change in the reflection
spectrum depends on both the nature of the inhomogeneity and the
strength of the incident X-rays. For the two-density models, where the
calculations could be distinguished by the original ionization
parameter $\xi$ of the uniform slab, the greatest impact on the \fe\
and \oeight\ Ly$\alpha$ EWs occurred when $\tau_{\mathrm{step}} \la 1$
for $\xi \la 1000$. For larger ionization parameters, steps deeper
into the gas could make an impact. This correlation is not expected to
continue indefinitely however. For example, if an atmosphere
had a drop in density at $\tau_{\mathrm{T}} \sim 3$ and it was
illuminated to the extent that gas was ionized down to this depth,
then the additional line emission or absorption caused by the
inhomogeneity would be smeared out by Compton scattering while escaping the
layer. Thus, a general conclusion seems to be that density
inhomogeneities must be within 2 Thomson depths of the surface to have
any impact on the reflection spectrum.

However, there must also be a lower-limit to the depth of any density
change in the reflecting medium for it to be effective in altering the
spectrum. In the extreme case of cold reflection (e.g., $\xi
\approx 10$) ionization effects are important only at $\tau_{\mathrm{T}} \leq
0.1$ from the surface. A density inhomogeniety at such a depth would
change such a small amount of gas that it would have a negligible effect on the
resulting spectrum.
 
The results of the two-density models also indicate that relatively
small changes in density (say, $\delta \rho/\rho \sim$few) can alter
the reprocessed emission. Again, this depends on the effective
ionization parameter of the layer, but, as was seen in
\S~\ref{sect:bubbles}, underdense regions beneath the surface of a
moderately irradiated atmosphere can result in a substantial 6.7~\kev\
iron line. But, if that region was underdense but a large amount, such
as 100--1000 times lower, then it will be ionized to such an extent
that it has very little effect on the outgoing spectrum. Similarly, an
overdense region below the surface may enhance the 6.4~\kev\ \fe\
line. Analogous arguments also apply to the soft X-ray emission lines
such as \oeight.

How realistic are these requirements on the density inhomogeneities?
It certainly appears plausible, even likely, that accretion flows
will naturally generate density contrasts greater than a few,
especially in the radiation dominated regime where the photon bubble
instability can enhance already existing clumpiness. What remains
unknown is if the heterogeneous nature of the flow exists to small
enough scales that sufficient inhomogeneity remains within a couple of
$\tau_{\mathrm{T}}$ from the disk photosphere. Only very high
resolution simulations of this region of accretion disks will be able
to definitively answer that question.

Assuming for the moment that the inhomogeneities discussed in this
paper occur in reality, what are the immediate observational
consequences? Perhaps the most interesting result is that the density
jumps can cause the \fe\ line to vary in a way that is
\emph{completely disconnected from the X-ray continuum}. One of the
most well known puzzling properties of the possible \fe\ `disklines'
is their lack of response to the variable continuum
\citep[e.g.][]{chi00}. The line flux typically does vary, but it is
often not directly correlated with the continuum
\citep{iwa96,wgy01,wwz01,ve01,mev03,iwa03}. One very striking example,
which was presented by \citet{pet02}, is an \textit{XMM-Newton}
observation of Mkn~841 that was split into two parts, separated by
about 15 hours. These authors found that the EW of the narrow \fe\
dropped by $\sim 2$ between the two observations while the continuum
changed by $\sim 10$--$20$\%. Although, we are not explicitly modeling
this source, Fig.~\ref{fig:ratiopanel} shows that density steps in the
surface can give rise to exactly such changes in the \fe\ EW,
completely independently of the X-ray continuum. A clear way to test
if this is the correct model would be to examine the changes of other
emission features in the spectrum, but this would require much more
sensitive data. The combination of density inhomogeneities with the
non-monotonic evolution of the \fe\ EW due to ionization
effects \citep{br02} may naturally result in a poor correlation between
the observed flux and the \fe\ EW. However, since the observed X-rays
must be averaged for many kiloseconds before spectral analysis, which
is longer than the timescale for a change in the density
inhomogeneities, as well as the the ionization/recombination
timescale, a relationship between the line and continuum may still be
uncovered, but at a weaker level than what may have been previously expected. 

If the density inhomogeneities beneath the photosphere are
relatively small in scale, which should be true for thin disks (where
$H < R$, and $H$ is the disk scale height and maximum size of any
density fluctuations), then any resulting enhancement or change to the \fe\
line would only occur to a narrow part of the overall line profile,
which is determined by the illuminating emissivity. Thus, the
appearance of narrow \fe\ lines, such as those recently inferred from
observations of NGC~3516 \citep{ttj02}, NGC~7314 \citep{yaq03}, and
Mrk~766 \citep*{tkr03} could conceivably originate from density
inhomogeneities in the disk.  

Variable \fe\ lines from turbulent accretion disks have also been
  considered by \citet{ar03}. These authors assumed the line
  emissivity was proportional to the local integrated magnetic stress
  in their numerical simulations, as opposed to our method of assuming
  a constant illumination and a variable density structure. In
  reality, a mixture of the two effects would be expected to be
  ongoing, which should be investigated in future work.

The strength of the soft X-ray emission lines are also affected by
density inhomogeneities. These lines are one of the dominant sources
of cooling for the X-ray heated gas when it reached $\la
10^6$~K. Thus, if overdense or underdense regions are introduced into
the layer it can alter the rapidity at which the gas cools, thereby
changing the emission features. Interestingly, these changes may occur
only in the soft band, with little effect at higher energies. Soft
X-ray lines, especially the Ly$\alpha$ lines of \oeight, \ion{N}{7}
and \ion{C}{6}, have received some attention recently with the claims
that they have been observed to be relativistically broadened in the
gratings spectra of some Seyfert~1s \citep{bra01,mas03,sak03}. The
exact strength of the lines have been the subject to some debate in
the literature \citep{lee01,brf02,tur03}, with the difficulties
arising because complex warm absorption features must be taken into
account in the spectral modeling. The results from the two-density models
show that density inhomogeneities can both enhance and diminish the
EWs of the soft X-ray lines.

The constant density fits to both the two-density models and the spectra
computed from the photon bubble structure resulted mostly in
reflection fractions much less than the true value of unity. If one
interprets $R$ as a measure of the solid angle subtended by the disk
as seen from the X-ray source, then $R < 1$ would indicate some form
of truncated accretion flow
\citep*[e.g.,][]{zds97,zds98,zds99,done99,esm00,gse03}. Correctly
interpreting a fitted value of $R$ is fraught with difficulty since
fitting neutral reflection models to ionized accretion disks will also
give $R < 1$ even if $R=1$ \citep{brf01,dn01}. Here, we have shown that
ionized disk models can produce erroneous $R$ values when there
are density inhomogeneities. It is therefore important that other
arguments be used \citep*[e.g.,][]{bdn03} before drawing conclusions
on the accretion geometry from a fitted value of $R$.

\section{Conclusions}
\label{sect:concl}
Our conclusions from this first paper in our study on reflection from
heterogeneous accretion disks can be summarized as follows:

\begin{enumerate}
\item The reflection spectrum from accretion disks can be altered by
  clumps or voids in the gas just beneath the photosphere. The effects
  range from negligible to large and depend on the structure of the
  inhomogeneity and the illuminating continuum.

\item The greatest effects on the reflected emission for most ionization
  parameters occurs when the density change is within $\sim 2$ Thomson
  depth of the surface.

\item The change in density does not have to be large, even an
 increase or a decrease by a factor of a few can significantly alter
 the spectrum from that of a constant density model.

\item Density inhomogeneities beneath the disk
  surface are a possible explanation for the apparent disconnectedness
  between the \fe\ line and the X-ray continuum.

\item The soft X-ray emission lines (below 1~\kev) are more sensitive to
  the presence of an inhomogeneity than the harder emission.

\item Fitting the spectra produced from models with inhomogeneities
  with constant density models shows that a mixture of ionization
  parameters are present, which often results in reflection fractions
  smaller than unity.

\item Constant density models may still be a good means of
  parameterizing reflection spectra, but only for energies $\ga
  2$~\kev. They will have much more difficulty with broadband spectra
  that cover the soft X-ray lines emitted from the disk. However, in
  practice this may only be a problem for sources with weak or
  non-existent warm absorbers. 

\end{enumerate}

In Paper 2, we continue our investigation by computing reflection
spectra from the predictions of 3-D radiation dominated accretion disk
simulations. Future work will also include a consideration of the
dynamical effects of the incident X-rays and multi-dimensional reflection.

\acknowledgments

DRB acknowledges financial support by the Natural Sciences and Engineering
Research Council of Canada, and thanks all the staff at KITP for their
hospitality during his visit. This research was supported in part by
the National Science Foundation under Grant No. PHY99-07949.

\clearpage

\begin{deluxetable}{cccccccccccccc}
\tabletypesize{\small}
\rotate
\tablewidth{0pt}
\tablecaption{\label{table:stepfits}Results of fitting a sample of two-density
  models with constant density reflection spectra.}
\tablehead{
 & & & &\multicolumn{4}{c}{0.2--12~\kev} & &
  &\multicolumn{4}{c}{2.0--10~\kev}\\
\colhead{$\xi_{\mathrm{model}}$} & \colhead{$N$} & \colhead{} &
\colhead{} & \colhead{$\Gamma$}
  & \colhead{$\xi$} & \colhead{$R$} & \colhead{$\chi^2$/d.o.f.} &
\colhead{}& \colhead{} &
  \colhead{$\Gamma$} & \colhead{$\xi$} & \colhead{$R$} &
  \colhead{$\chi^2$/d.o.f.}
}
\startdata
125 & 10$^{-23}$ & & & 2.021$^{+0.004}_{-0.003}$ & 147$\pm 3$ &
0.79$^{+0.01}_{-0.02}$ & 2672/2358
& & & 2.03$^{+0.01}_{-0.02}$ & 466$^{+38}_{-42}$ & 1.2$\pm 0.2$ & 1684/1595\\
250 & 10$^{-23}$ & & & 2.026$^{+0.002}_{-0.004}$ & 190$\pm 3$ &
0.92$^{+0.02}_{-0.01}$ & 4265/2358
& & & 2.03$\pm 0.01$ & 327$^{+40}_{-42}$ & 1.1$\pm 0.1$ & 1590/1595\\
500 & $5\times 10^{-24}$ & & & 2.0 & 631 & 0.654 & 7833/2358
& & & 2.02$\pm 0.01$ & 507$^{+20}_{-33}$ & 0.88$^{+0.08}_{-0.09}$ & 1638/1595\\
1000 & $2\times 10^{-24}$ & & & 2.000$^{+0.004}_{-0.001}$ &
1244$^{+12}_{-17}$ & 0.75$\pm 0.02$ & 2994/2358 & & & 2.03$\pm 0.01$ &
1282$^{+170}_{-181}$ & 0.81$\pm 0.05$ & 1639/1595\\ 
\enddata
\tablecomments{The step parameters for the two-density model spectra are
  (a) $\xi=125$, $f_{\rho}=0.1$, $\tau_{\mathrm{step}}=0.1$, $k=2$;
  (b) $\xi=250$, $f_{\rho}=10$, $\tau_{\mathrm{step}}=0.5$, $k=0.2$;
  (c) $\xi=500$, $f_{\rho}=10$, $\tau_{\mathrm{step}}=0.5$, $k=1$; (d)
  $\xi=1000$, $f_{\rho}=0.001$, $\tau_{\mathrm{step}}=0.5$, $k=2$. $N$
  is the normalization of the two-density models in \textsc{xspec}. $R$ is the
  reflection fraction defined as
  total=incident+$R\times$reflected. $\xi$ is in units of
  erg~cm~s$^{-1}$. The error-bars are the 2$\sigma$ uncertainty for
  the parameter of interest. The simulated data were constructed using
  the \textit{XMM-Newton} response matrix \texttt{epn\_sw20\_sdY9.rmf}
  and assumed an exposure time of 40~ks.}
\end{deluxetable}

\clearpage

\begin{deluxetable}{cccccccccccccc}
\tabletypesize{\small}
\rotate
\tablewidth{0pt}
\tablecaption{\label{table:bubblefits}Results of fitting the
  photon bubble reflection spectra with constant density models.}
\tablehead{
 & & & &\multicolumn{4}{c}{0.2--12~\kev} & &
  &\multicolumn{4}{c}{2.0--10~\kev}\\
\colhead{$F_{\mathrm{X}}$} & \colhead{$N$} & \colhead{} &
\colhead{} & \colhead{$\Gamma$}
  & \colhead{$\xi$} & \colhead{$R$} & \colhead{$\chi^2$/d.o.f.} &
\colhead{}& \colhead{} &
  \colhead{$\Gamma$} & \colhead{$\xi$} & \colhead{$R$} &
  \colhead{$\chi^2$/d.o.f.}
}
\startdata
10$^{13}$ & $2\times 10^{-20}$ & & & 1.988$^{+0.030}_{-0.001}$ &
561$^{+1}_{-9}$ & 1.01$\pm 0.01$ & 4679/2358
& & & 2.04$\pm 0.01$ & 601$^{+42}_{-40}$ & 0.96$^{+0.11}_{-0.10}$ & 1632/1595\\
$2.5\times 10^{13}$ & 10$^{-20}$ & & & 1.950 &
565 & 0.798 & 5821/2358
& & & 2.03$\pm 0.01$ & 1297$^{+122}_{-92}$ & 0.73$^{+0.13}_{-0.04}$ &
1695/1595\\ 
$5\times 10^{13}$ & $5\times 10^{-21}$ & & & 1.970 & 593 & 0.926 & 6651/2358
& & & 2.00$\pm 0.01$ & 1340$^{+488}_{-238}$ & 0.52$\pm 0.03$ & 1561/1595\\
10$^{14}$ & $2\times 10^{-21}$ & & & 2.029 & 491 & 0.611 & 5995/2358 &
& & 1.987$^{+0.007}_{-0.008}$ & 2679$^{+334}_{-297}$ & 0.46$\pm 0.03$
  & 1627/1595\\ 
\enddata
\tablecomments{$N$ is the normalization of the photon bubble models in
  \textsc{xspec}. $R$ is the reflection fraction defined as
  total=incident+$R\times$reflected. $\xi$ is in units of
  erg~cm~s$^{-1}$. $F_{\mathrm{X}}$ is in units of
  erg~cm$^{-2}$~s$^{-1}$. The error-bars are the 2$\sigma$
  uncertainty for the parameter of interest. The simulated data were
  constructed using the \textit{XMM-Newton} response matrix
  \texttt{epn\_sw20\_sdY9.rmf} and assumed an exposure time of
  40~ks.}
\end{deluxetable}

\clearpage


\clearpage 

\clearpage 

\clearpage

\clearpage

\clearpage

\begin{figure}
\figurenum{\ref{fig:bubblespect} \emph{cont}}
\plottwo{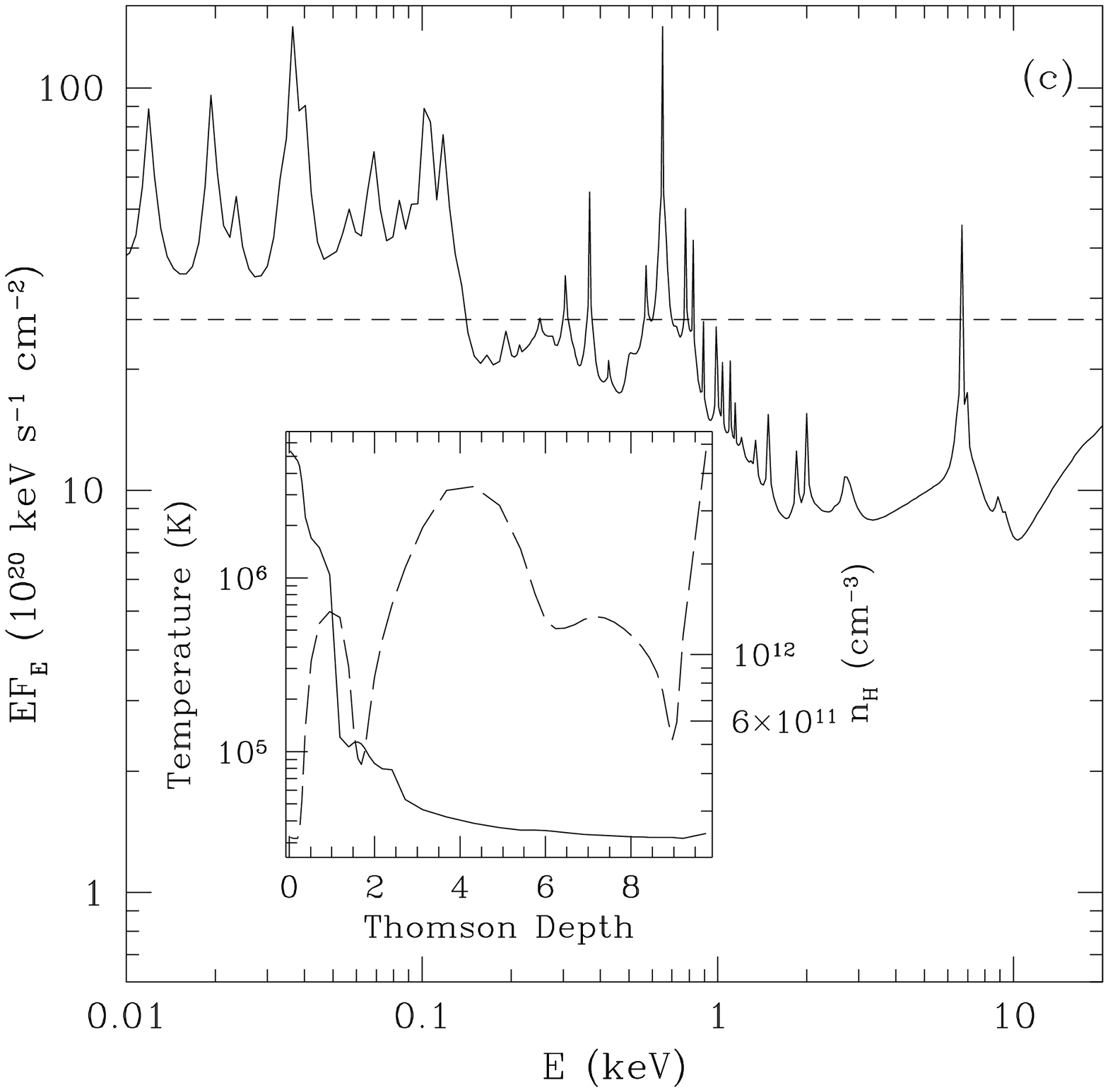}{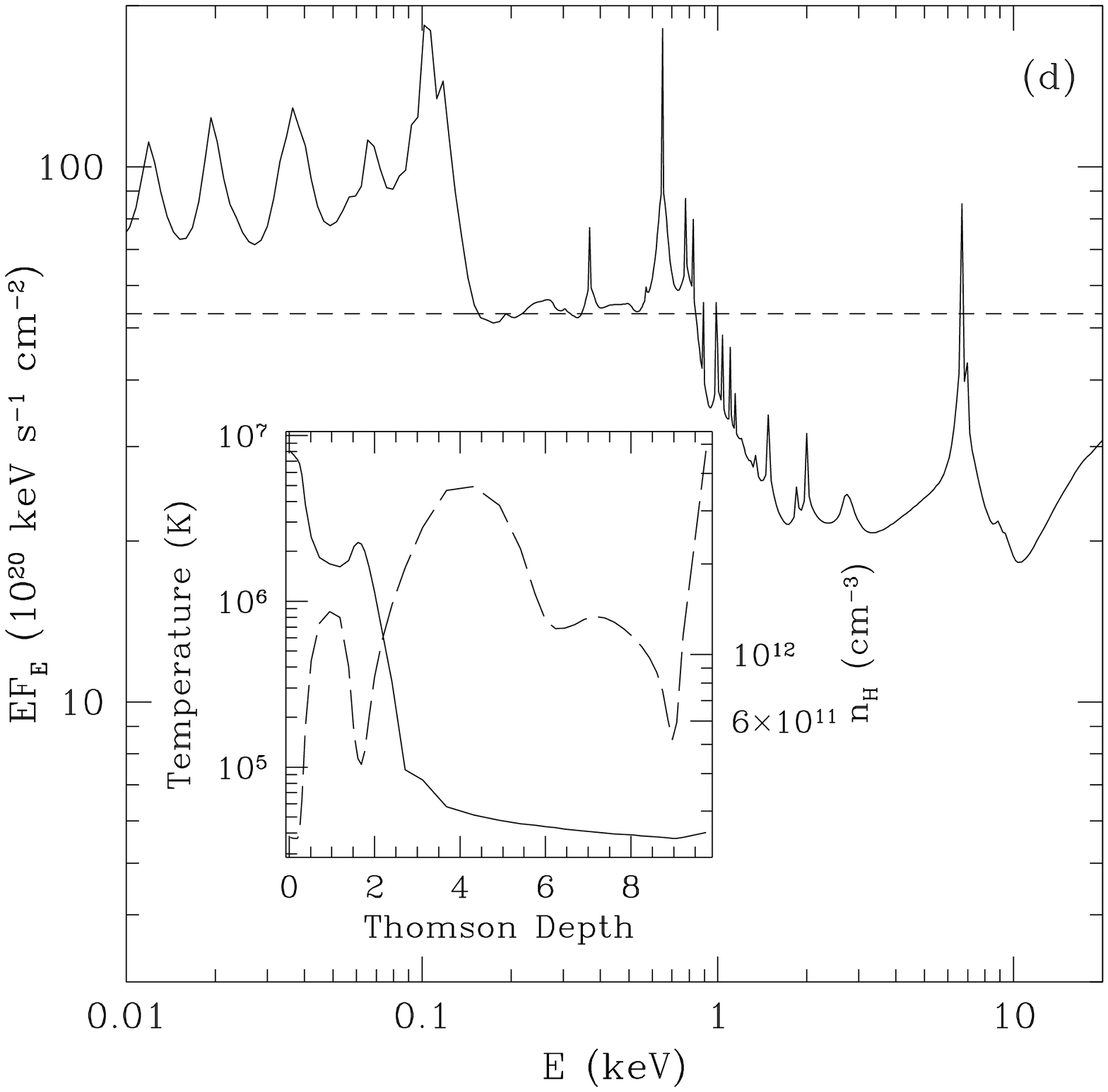}
\caption{(c) $F_{\mathrm{X}}=5\times
  10^{13}$~erg~cm$^{-2}$~s$^{-1}$, (d)
  $F_{\mathrm{X}}=10^{14}$~erg~cm$^{-2}$~s$^{-1}$.}
\end{figure}

\clearpage

\end{document}